\documentclass[namedreferences]{solarphysics}
\usepackage[hyperref,optionalrh,solaromanenum]{spr-sola-addons} 
\usepackage[pdftex]{graphicx}      
\usepackage{color}                 

\chardef\us=`\_

\begin{document}
\begin{article}

\begin{opening}

\title{The revised Brussels-Locarno Sunspot Number (1981-2015)}

\author[addressref={afil1},corref,email={}]{\inits{F.}\fnm{Fr\'{e}d\'{e}ric}~\lnm{Clette}}
\author[addressref={afil1},corref,email={}]{\inits{L.}\fnm{Laure}~\lnm{Lef\`{e}vre}}
\author[addressref={afil2},corref,email={}]{\inits{M.}\fnm{Marco}~\lnm{Cagnotti}}
\author[addressref={afil2},corref,email={}]{\inits{S.}\fnm{Sergio}~\lnm{Cortesi}}
\author[addressref={afil3},corref,email={}]{\inits{A.}\fnm{Andreas}~\lnm{Bulling}}
 
\runningauthor{Clette et al.}
\runningtitle{The 1981--2015 Brussels--Locarno Sunspot Number}

\address[id={afil1}]{Royal Observatory of Belgium, 3 avenue Circulaire, 1180 Brussels, Belgium}
\address[id={afil2}]{Specola Solare Ticinese, Locarno, Switzerland}
\address[id={afil3}]{SONNE, Vereinigung der Sternfreunde, Germany}

\begin{abstract}
In 1981, the production of the international Sunspot Number moved from the Z\"{u}rich Observatory to the Royal Observatory of Belgium, with a new pilot station, the Specola Solare Ticinese Observatory in Locarno. This marked a very important transition in the history of the Sunspot Number. Those recent decades are particularly important as they provide the link between recent modern solar indices and fluxes and the entire Sunspot Number series extending back to the $18^{th}$ century. However, large variations have been recently identified in the scale of the Sunspot Number between 1981 and the present.

Here, we refine the determination of those recent inhomogeneities by reconstructing a new average Sunspot Number series $\rm S_N$ from a subset of long-duration stations between 1981 and 2015. We also extend this reconstruction by gathering long-time series from 35 stations over 1945-2015, thus straddling the critical 1981 transition. In both reconstructions, we also derive a parallel Group Number series $\rm G_N$ built by the same method from exactly the same data set.

Our results confirm the variable trends associated with drifts of the Locarno pilot station, which start only in 1983. We also verify the scale of the resulting 1981-2015 correction factor relative to the preceding period 1945--1980, which leads to a fully uniform $\rm S_N$ series over the entire 1945--2015 interval. By comparing the new $\rm S_N$ and $\rm G_N$ series, we find that a constant quadratic relation exists between those two indices. This proxy relation leads to a fully constant and cycle-independent $\rm S_N/G_N$ ratio over cycles 19 to 23, with the exception of cycle 24 where it drops to a lower value. Several comparisons with other indices show a very good agreement between our reconstructed $\rm G_N$ and the new ``backbone'' Group Number \citep{Clette_etal_2014} but reveal inhomogeneities in the original Group Number as well as the $\rm F_{10.7}$ radio flux and the American sunspot number $\rm R_a$ over the period 1945--2015.

We conclude on the new potential diagnostics that can emerge from the new re-calibrated Sunspot Number. This analysis also opens the way to the implementation of a more advanced method for producing the Sunspot Number in the future, based on various conclusions from the present recalibration. In particular, we identify the existence of distinct subsets of observing stations sharing very similar personal k factors, which may be a key element for building a future multi-station reference in place of the past single pilot station.
\end{abstract}


\end{opening}

\section{Introduction} \label{S-Intro} 
Although it is presented as a single continuous time series, the Sunspot Number (hereafter {\bf SN}) series was actually constructed in a succession of segments, each corresponding to different observers, instruments and methods. An overview of the SN history is given by \citet{Clette_etal_2014}. Most of this construction was carried out over 181 years (1849-1980) at the federal Observatory of Z\"urich in the footsteps of Rudolf Wolf, its first Director, who introduced the well-know definition of the {\bf Wolf number}, condensing one observation by one observer into a single global measure of sunspot activity:
\begin{equation}
W_S = k \times (10\,N_g + N_s)    \label{EQ_wolfnum}
\end{equation}
where $N_g$ is the number of sunspot groups, $N_s$ the total number of spots and  $k$ a scaling coefficient specific to each observer.

Therefore, the transfer of the World Data Center (WDC) for the Sunspot Number from Z\"{u}rich to Brussels, in 1981, marks one of the most drastic changes in the production of the SN. Indeed, when taking over the WDC at the Royal Observatory in Brussels, Andr\'{e} Koeckelenbergh introduced refinements in the computing method, fully computerizing the calculation of the SN formerly done largely by hand \citep{Berghmans_etal_2006, Clette_etal_2007}. However, although the new calculation included all data from all stations and the network of observing stations was rapidly expanded to reach about 80 stations, the long-term scale of the resulting SN was still resting on a single pilot station, in the continuity of the original principles used in Z\"{u}rich. Therefore, the prime change occurring in 1981 was the replacement of the Z\"{u}rich Observatory by the Specola Solare Observatory in Locarno. Indeed, the Z\"{u}rich Observatory was closed, thus terminating its long reference series (cf. the paper by J. Stenflo in this issue, \citep{Stenflo_2015})

The primary observer at the Specola, Sergio Cortesi, had been fully trained by Max Waldmeier, last Director of the Z\"{u}rich Observatory and had a track record of consistent observations over more than 25 years (starting in 1958). This thus offered the best guaranties of a full continuity across the 1981 transition. Unfortunately, \citet{Clette_etal_2014} recently found that soon after 1980, the Wolf numbers from the Specola Observatory started to drift, first becoming too high over cycle 21 and then declining until the end of cycle 23, before reversing again over the past 8 years. Here, we will not attempt to diagnose the causes of those variations. This investigation has been started but may take time, as counts made at the eyepiece leave limited information. On the other hand, given the abundance of data from many observers over the last 35 years, we will instead exploit this wealth of parallel observations.

Fixing the defects in this last segment of the SN series is particularly important as it is precisely over the last decades that most of the many modern solar data and indices started to be measured, giving a detailed view of multiple processes driving solar activity or driven by it. Therefore, the last link in the SN chain provides the key for transposing modern knowledge to epochs in the remote past when the Sun went trough different regimes of activity, including the Grand Minima (solar irradiance, Sun-climate connection). This proved particularly true over recent years, as the current weak cycle 24 brings us back to activity levels not encountered since more than one century, i.e. never recorded with most of our current observing techniques. 

In this paper, we first describe the set of data assembled to build a new homogeneous SN series. We then derive a first accurate SN over the interval 1981-2015 using the large set of stations in the SILSO network. As this series must be brought to the same scale as the Z\"{u}rich SN before 1981, we then build a longer SN series from 1944 to 2015 using a smaller set of long-duration stations. The main parallel long-term series at our disposal is the Group Number \citep{Hoyt-Schatten_1998a, Hoyt-Schatten_1998b}. However, various inhomogeneities have been found in this series, including a $\sim10 \% $ scale change around 1975 \citep{Lockwood_etal_2014, Clette_etal_2014}. In order to correct this defect and be able to make reliable comparisons with the corrected SN, we thus build a new group number using the same set of stations. We then investigate the relation between those two sunspot-based indices, now free from artifacts, and we make a few comparisons with other standard solar indices. We finally summarize the final corrections and some new insights on other solar measurements that emerge thanks to the greatly improved stability of the new SN series.  
 
\section{The data} \label{S-Data} 
For the core ``Brussels--Locarno'' period, we can fully exploit the SILSO database where more than 560\,000 observations from 277 stations are preserved. About two thirds of the station have been observing for less than 11 year. They bring only limited information about the long term stability of the SN as they can at best cover a fraction of two successive cycles. As such, they were mostly dropped from this analysis.

The remaining long-duration stations were then submitted to a selection based on stability criteria. By a simple visual inspection of data plots, we first eliminated stations with obvious defects found only for one single station: large jumps or dips, disproportionate trends or strongly distorted solar cycles. A second more quantitative selection was applied iteratively as part of the construction of the new average SN series. The stations were ranked according to the degree of linear correlation to the reference average SN and according to the dispersion of the residuals (standard deviation), i.e. according to both their overall homogeneity and their precision.

\begin{figure} 
	\centerline{\includegraphics[width=1.0\textwidth,clip=true,trim= 15 0 5 0,clip=true]{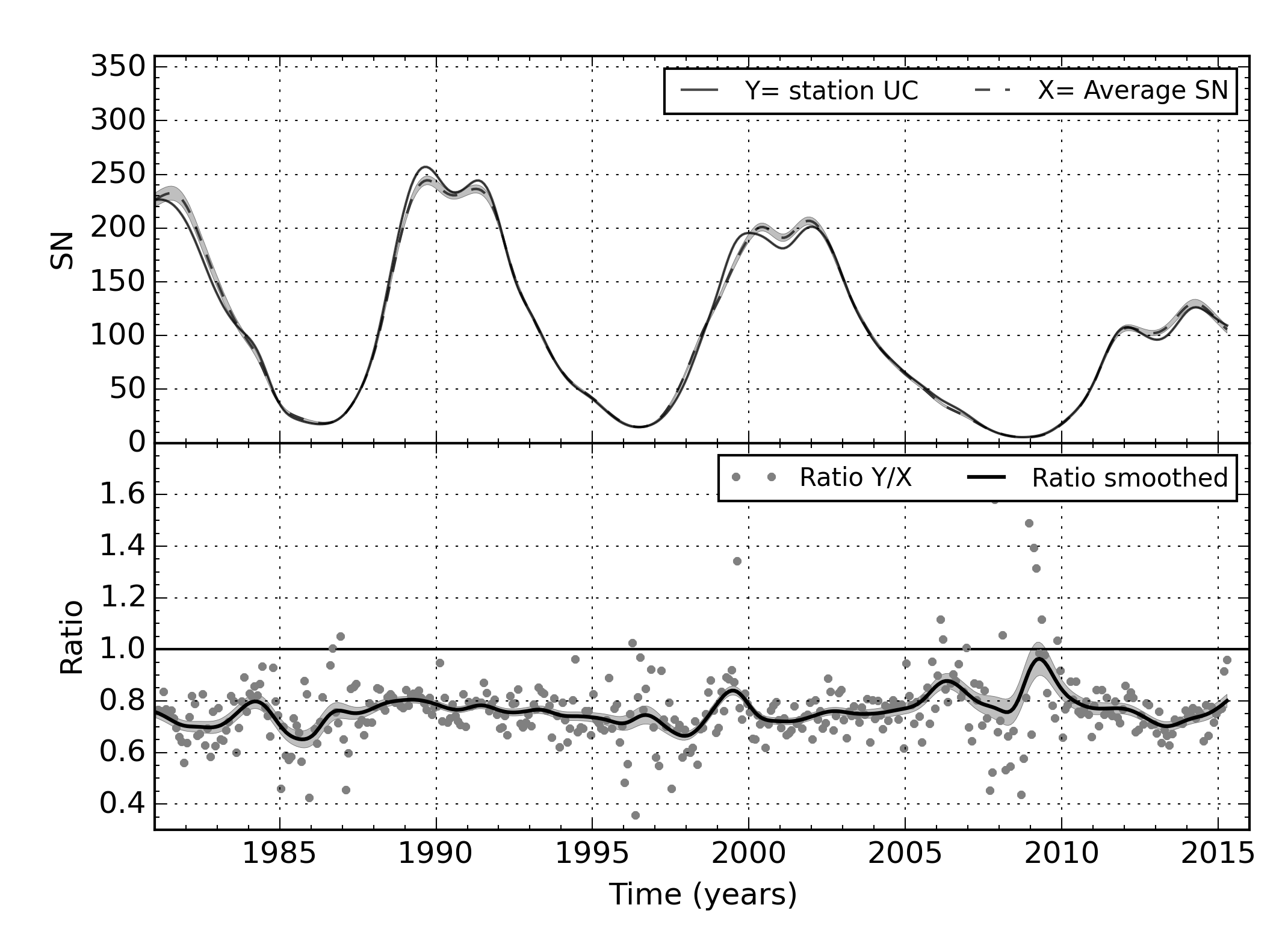}}
	\caption{Example of a stable station (Uccle station, ROB, Brussels). The top panel shows the raw monthly mean Wolf numbers from Uccle, smoothed by a 12-month Gaussian running mean (solid line) compared to the average SN series reconstructed as described later in this paper (dashed line). The two series were brought to the same scale by applying the mean k ratio between the original series. The lower panel shows the ratio between those two series (dots: monthly mean ratio; solid line: 12-month Gaussian smoothing). An estimate of the standard error on the mean ratio is shown by the grey shading. Except for larger fluctuations simply due to low SN values near cycle minima, the ratio is uniform over the whole 35--year interval.} 
	\label{Fig-exstagood}
\end{figure}

\begin{figure} 
	\centerline{\includegraphics[width=1.0\textwidth,clip=true,trim= 15 0 5 0,clip=true]{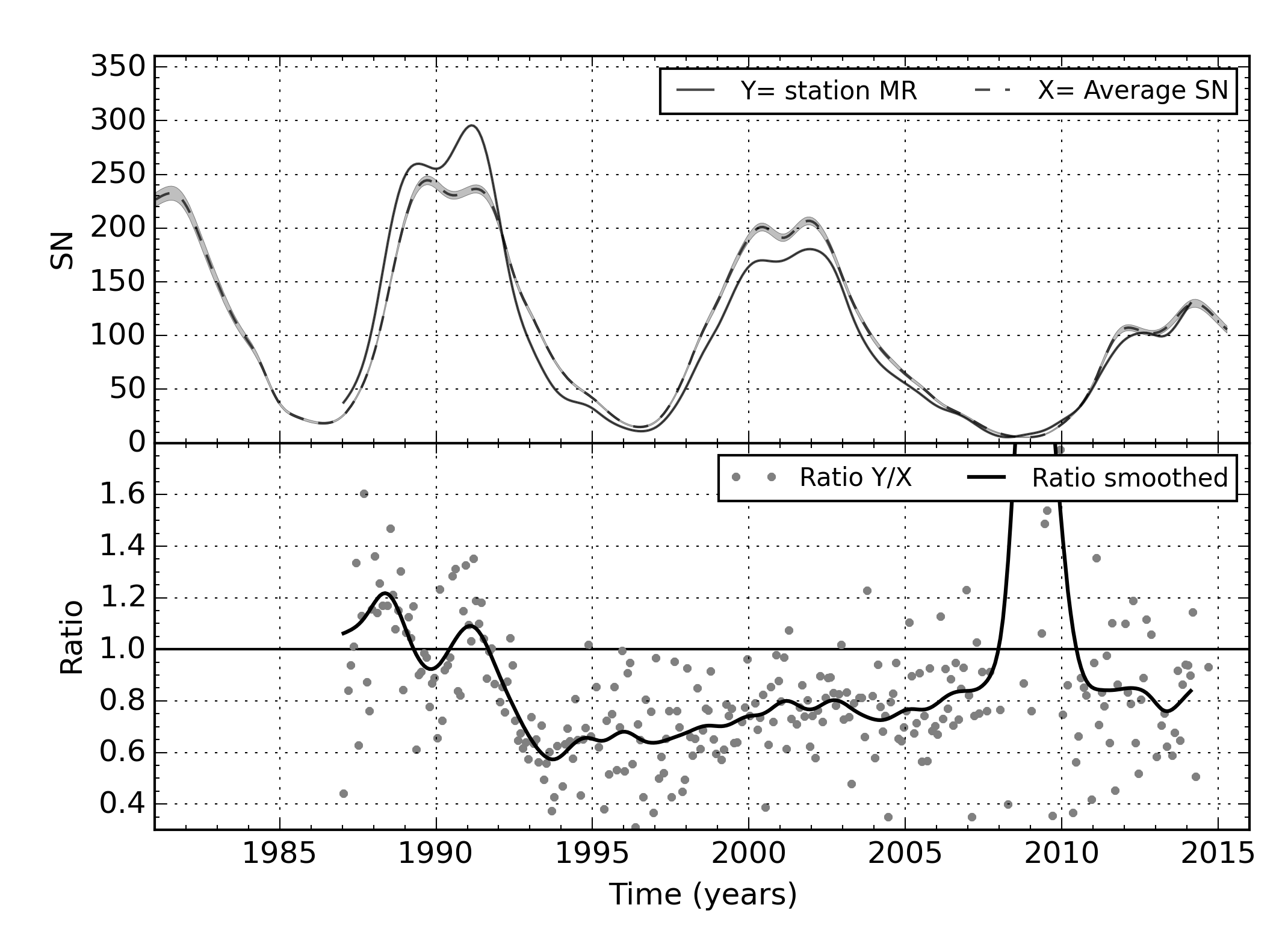}}
	\caption{Example of an unstable station (Mira observatory, Grimbergen, Belgium), using the same representation as in figure \ref{Fig-exstagood}. This station shows a much larger dispersion in its monthly values, as well as marked variable trends. This is a public observatory with a changing pool of volunteer observers, which explains the long-term inhomogeneity of the series.}  
	\label{Fig-exstabad}
\end{figure}

Various sets of stations were selected, including either few very good stations or adding more stations but accepting lower stability. About 80 stations could be included in this analysis, all giving very high linear correlations (coefficient of determination $r^2$ from 0.92 to 0.98). As an illustration, figures \ref{Fig-exstagood} and \ref{Fig-exstabad} give examples of a stable station (Uccle) and of a strongly inhomogeneous station (Mira Observatory), which was excluded from our analysis.

For our extended analysis, we looked for long-duration stations that were active over the interval 1945-2015, thus straddling the 1981 transition. For the period before 1981, the WDC-SILSO could recover data from paper archives provided by the Specola Observatory. Those archives contained original observing reports submitted by auxiliary stations to the Z\"{u}rich Observatory mostly between 1950 and 1980. Unfortunately, some years are missing is this archive, most notably 1980, which falls just before the transition. In order to fill those gaps and also to increase our sample of stations, we looked for additional stations not included in the Z\"{u}rich  archives. We benefited from newly recovered or updated data from several sources:
\begin{itemize}
	\item the fully digitized series from Kislovodsk Observatory (courtesy A. Tlatov)
	\item the fully updated and digitized series from the Kanzelh\"{o}he Observatory (courtesy W. Po\"{e}tzi)
	\item digitized counts from H.\,Luft and H.\,Koyama (courtesy L. Svalgaard)
	\item newly recovered data from 8 German observers (SONNE, A.Bulling). Although those series only cover the interval 1972-1990, they bring important information around 1981.
	\item a newly recovered very long series by T.\,Cragg (1945-2008, AAVSO, courtesy R. Howe)
\end{itemize}
As some of those observers also sent their observations to Z\"{u}rich or Brussels, the integration of those series in this analysis allowed to fill some gaps or to fix some anomalous values in the source archives. The same selection procedure as for the 1981-2015 was applied to those stations, leaving 17 useful series over the interval 1944-2015 (Table \ref{tab_data1945}). Here again, as a verification,  we formed several sets after ranking the stations according to their quality. 

\begin{table}
\caption{List of long-duration stations used for the extended analysis 1945--2015, ranked by decreasing number of observed months. For each station, we give its k personal coefficient, the error on k,  the number of observed months (NM) and the coefficient of determination $r^2$ (quality of fit). The most stable stations used in the final reconstruction are marked by a ``$\ast$'' in the last column.}
\label{tab_data1945}

\begin{tabular}{llrrrrc}     
\hline
ID     & Station                 & \multicolumn{1}{c}{k} &\multicolumn{1}{c}{$\sigma_k$}
       & \multicolumn{1}{c}{NM} & \multicolumn{1}{c}{$r^2$} &  \\
\hline
KZm    & Kanzelh\"{o}he Obs., Austria        &  1.230 & 0.005 & 835 & 0.958 & $\ast$ \\
KS2    & Kislovodsk Obs., Russia             &  1.298 & 0.005 & 736 & 0.972 & $\ast$ \\
CRA    & T.A.Cragg, AAVSO, USA -- Australia  &  1.601 & 0.007 & 724 & 0.966 & $\ast$ \\
LO     & Specola Obs., Locarno, Switzerland  &  1.007 & 0.004 & 674 & 0.976 & $\ast$ \\
UC     & Uccle Obs., Brussels, Belgium       &  1.352 & 0.006 & 673 & 0.964 & $\ast$ \\
CA     & Catania Obs., Italy                 &  1.286 & 0.008 & 615 & 0.928 &   \\
KOm    & H.Koyama, Japan                     &  1.267 & 0.006 & 587 & 0.968 & $\ast$ \\
SK     & Skalnate Obs., Slovakia             &  1.387 & 0.009 & 565 & 0.943 & $\ast$ \\
KH     & Kandilli Obs., Turkey               &  1.405 & 0.007 & 559 & 0.957 & $\ast$ \\
BN-S   & WFS, Berlin, SONNE,y                &  1.025 & 0.009 & 547 & 0.900 &   \\
FU     & K.Fujimori, Nagano, Japan           &  1.220 & 0.005 & 543 & 0.972 & $\ast$ \\
SC-S   & W.Schulze, SONNE, Germany           &  1.392 & 0.013 & 543 & 0.891 &   \\
QU     & Quezon, Philippines                 &  1.862 & 0.013 & 538 & 0.925 &   \\
HD-S   & R.Hedewig, SONNE, Germany           &  1.439 & 0.011 & 513 & 0.931 & $\ast$ \\
SM     & San Miguel Obs., Argentina          &  1.081 & 0.010 & 510 & 0.891 &   \\
LFm    & H. Luft, New York, USA              &  1.431 & 0.008 & 473 & 0.959 & $\ast$ \\
HU     & Hurbanovo Obs., Slovakia            &  1.331 & 0.010 & 463 & 0.934 &   \\
PO     & Postdam Obs., Germany               &  1.326 & 0.012 & 443 & 0.908 &   \\
MO     & E. Mochizuki, Urawa, Saitama, Japan &  1.149 & 0.006 & 428 & 0.972 & $\ast$ \\
SA     & Santiago, Chile                     &  1.741 & 0.019 & 382 & 0.893 &   \\
A3     & Elias, Athens, Greece               &  1.241 & 0.012 & 311 & 0.929 &   \\
BR-80S & H.J.Bruns (80mm), SONNE, Germany    &  1.019 & 0.009 & 278 & 0.956 & $\ast$ \\
RO     & Roma Obs., Italy                    &  1.727 & 0.019 & 277 & 0.915 &   \\
HP-S   & W.Hoffmann, SONNE, Germany          &  1.545 & 0.015 & 232 & 0.941 &   \\
MA     & Manila, Philippines                 &  1.264 & 0.009 & 200 & 0.976 & $\ast$ \\
LK     & Looks, Chile                        &  1.581 & 0.015 & 197 & 0.951 & $\ast$ \\
GU-S   & R.Guenther, SONNE, Germany          &  2.083 & 0.025 & 187 & 0.929 &   \\
HT-S   & S.Hammerschmidt, SONNE, Germany     &  1.567 & 0.018 & 176 & 0.961 &   \\
TR     & Trieste Obs., Italy                 &  1.718 & 0.017 & 152 & 0.945 &   \\
GA-S   & C.-D.Gahsche, SONNE, Germany        &  1.573 & 0.038 & 147 & 0.800 &   \\
FR-S   & Froebrich, SONNE, Germany           &  1.002 & 0.013 & 146 & 0.949 & $\ast$ \\
AT     & Eugenides, Athens, Greece           &  1.117 & 0.018 & 142 & 0.878 &   \\
AN     & Ankara Obs., Turkey                 &  1.761 & 0.020 & 137 & 0.950 & $\ast$ \\
BK-S   & Beck, SONNE, Germany                &  0.764 & 0.025 &  70 & 0.744 &   \\
MY-S   & Mey, SONNE, Germany                 &  1.869 & 0.043 &  62 & 0.698 &   \\
\hline
\end{tabular}
\end{table}

For each set, we formed three data series with different time resolutions: daily values, monthly means and yearly means. The analyses were applied to all three series, leading to almost identical results within the uncertainties. Here, we mainly present results from the monthly and yearly data, as they give the clearest view of the long-term variations and inhomogeneities. 
For the readability of plots shown in this paper, the results from monthly data were smoothed by a Gaussian smoothing function with a width at half maximum of 12 months.

As virtually all series provided separate group and spot counts, we formed parallel series of only the group count, which will allow us to build a GN series based on exactly the same stations by the same method.
 
\section{The method: a new strategy} \label{S-Method}
In a first approach (Clette et al. 2014), we started from the daily values of the personal k coefficient for each individual stations, defined as: $k= S_N / W_S $, the ratio between the original Sunspot Number $S_N$ and the raw Wolf number $W_S$ observed by the station on the same day. By averaging all k values from all station, we finally built a global k time series actually representing the variations of the common reference versus the entire SILSO network. The advantage of this method is to deliver directly a k correction factor that can be applied to the original SN, or more precisely to the Wolf Numbers from the Specola-Locarno pilot station and that shows directly any deviation from a constant ratio. 

However, as the mean of a ratio is different from the ratio of the means, this k correction can be slightly biased, in particular when the activity is low ($\rm W_S$ values in the denominator close to 0). Moreover, the method does not produce directly a new corrected SN series and the overall absolute scale of the k correction factor is difficult to establish. Indeed, this scale is relative and must be attached to the scale of the SN data before 1981. Moreover, as the SN before 1981 is mostly the Wolf number of a single station, there is no equivalent k factor available.

\subsection{Multi-station average Sunspot and Group Numbers} \label{SS-Multista}
Therefore, for this new analysis, our strategy was to work directly with the Wolf Numbers from all stations and then build a network-wide average SN series. As this reconstruction can be done with data both before and after the 1981 transition, it will thus allow matching the scales before and after 1981. The new reconstructed SN series can then either be substituted to the original SN or more indirectly, the k ratio between the two master series can be converted into monthly or yearly mean correction factors to be applied to original data. However, as the sample of stations is much larger after 1981, we split the analysis in two steps. First, we use a short 1981-2015 data set exploiting the whole set of data in the SILSO database in order to reach the highest possible accuracy over that interval. Then, we treat a longer 1945--2015 time interval but using only stations that were active both before and after the 1981 transition, i.e. avoiding a discontinuity in our sample of stations just in 1981. It should be noted that given the bounded time interval covered by the data archives, the number of simultaneous data is maximum in the middle of the interval and drops off slightly at both ends, typically by about 50\% at the outer limits. For the long series, this maximum sample falls quite appropriately around the critical 1981 transition. 

\subsection{2-step iteration} \label{SS-Iteration}
The individual series of all stations must be brought to a common scale before the values are averaged in order to give them equal weights. Indeed, we observed that the stability and precision of a station are not related to the fact that this station counts more or less spots on average compared to other stations. One of the stations must be chosen to act as the initial scaling reference. Naturally, for this reference, we picked one station spanning the longest possible period and with a good stability. In practice, the reference station and all other stations don't have observations for all days and they don't cover exactly the same time intervals. The reference series may also contain temporary scale fluctuations. As a consequence, the resulting k coefficient will be less accurate for stations with a limited number of matching daily data with the reference station. This is true in particular for stations only active at both ends of the analyzed period.

Therefore, we proceeded in two steps. We build a first average series using a single reference station. Then, we use this first reconstructed average SN or GN series as the scaling reference to determine the final k coefficients for all stations. We can then build a final average SN or GN series by averaging the normalized series scaled with these optimized k coefficients. The result is thus more immune to defects in a single reference series and reference values are available on all days for the k normalization. The overall principle is thus similar to the ``backbone method'' introduced by Svalgaard \citep[see][]{Clette_etal_2014} for the GN reconstruction, but with this 2-step process. This optimization is applicable in this case thanks to the fact that at any given time in the interval, data from many stations are available and thus the averaging can bring a significant gain in accuracy.

\section{Reconstructed Sunspot Number 1981--2015} \label{S-SN1981}
\begin{figure} 
	\centerline{\includegraphics[width=0.95\textwidth,clip=true,trim= 15 0 5 0,clip=true]{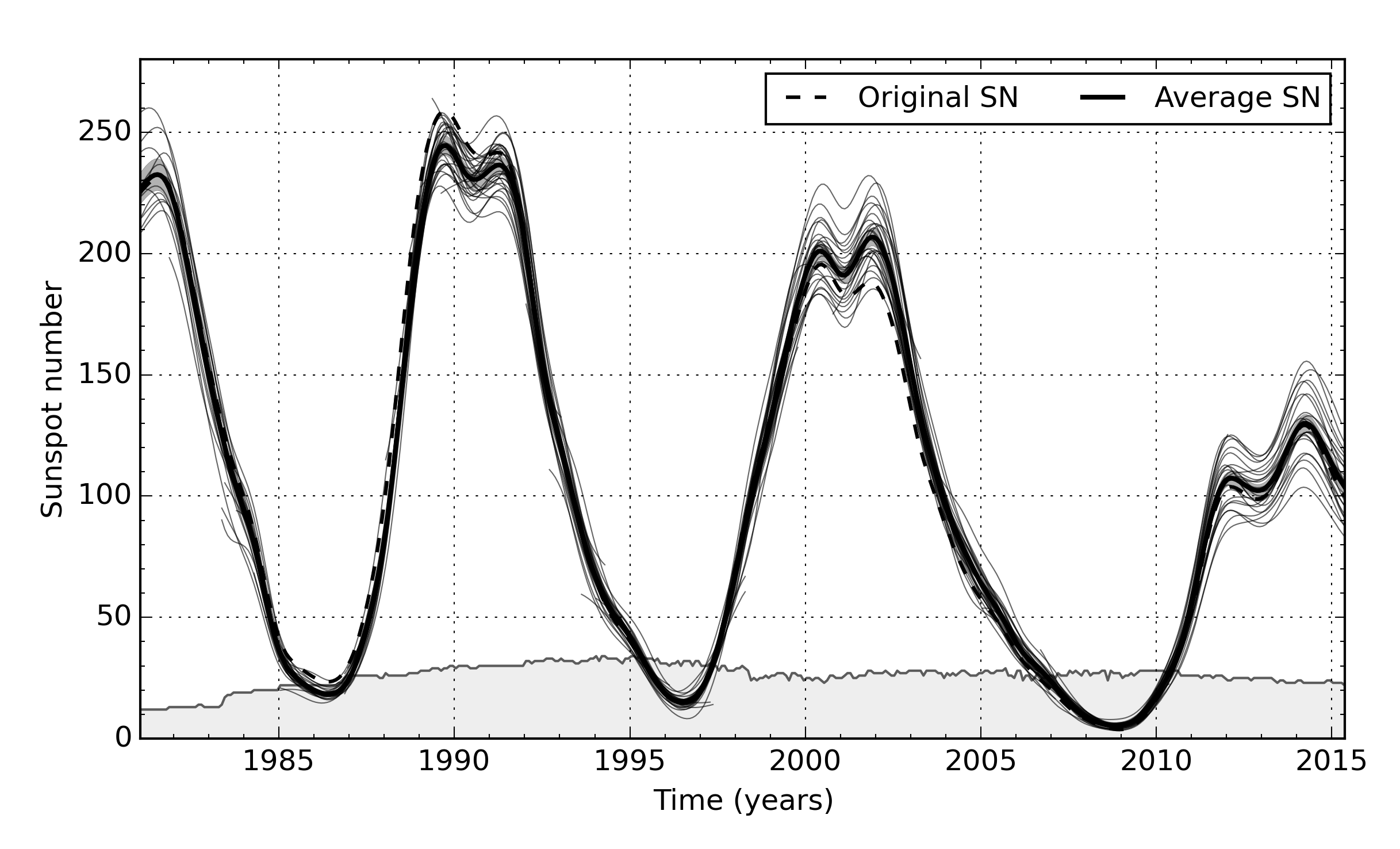}}
	\caption{Reconstructed average SN (thick solid curve) obtained by an average over 42 stations, normalized on the Specola--Locarno station over the entire interval 1981-2015. In order to better show the long-term trends, we applied a 12-month Gaussian smoothing to the monthly means. The thin curves show the individual series of all stations included in the calculation. The dashed curve shows the original SN series (scale divided by 0.6), while the gray shading around the average SN curve gives the standard error on the reconstructed monthly means. The shaded curve at the bottom gives the number of stations included in each monthly mean, ranging from 12 in the early 1980s up to 35 in the middle of the time interval.} 
	\label{Fig-avgsn}
\end{figure}

For the 1981-2015, we have at our disposal from 70 to 100 stations active at any given time, of which 28 stations delivered data for more than 30 years and form the core ``backbone'' of this analysis. Figure \ref{Fig-avgsn} shows the average SN series (monthly means) calculated from a base set of 42 stations together with the number of stations used for each monthly values. The selection process described in section \ref{S-Data} led to the elimination of several important professional observatories (e.g. Catania, Ebro, Mitaka), which proved to suffer from large trends or erratic jumps. Those defects probably result from changes of observer or equipment and will require further investigation with the help of those observatories. All SOON stations (Holoman, Learmonth, San Vito) give overestimated SN values during solar minima, strongly deviating from all other stations, in particular during the cycle 23-24 minimum. 

\begin{figure} 
	\centerline{\includegraphics[width=1.0\textwidth,clip=true,trim= 15 0 5 0,clip=true]{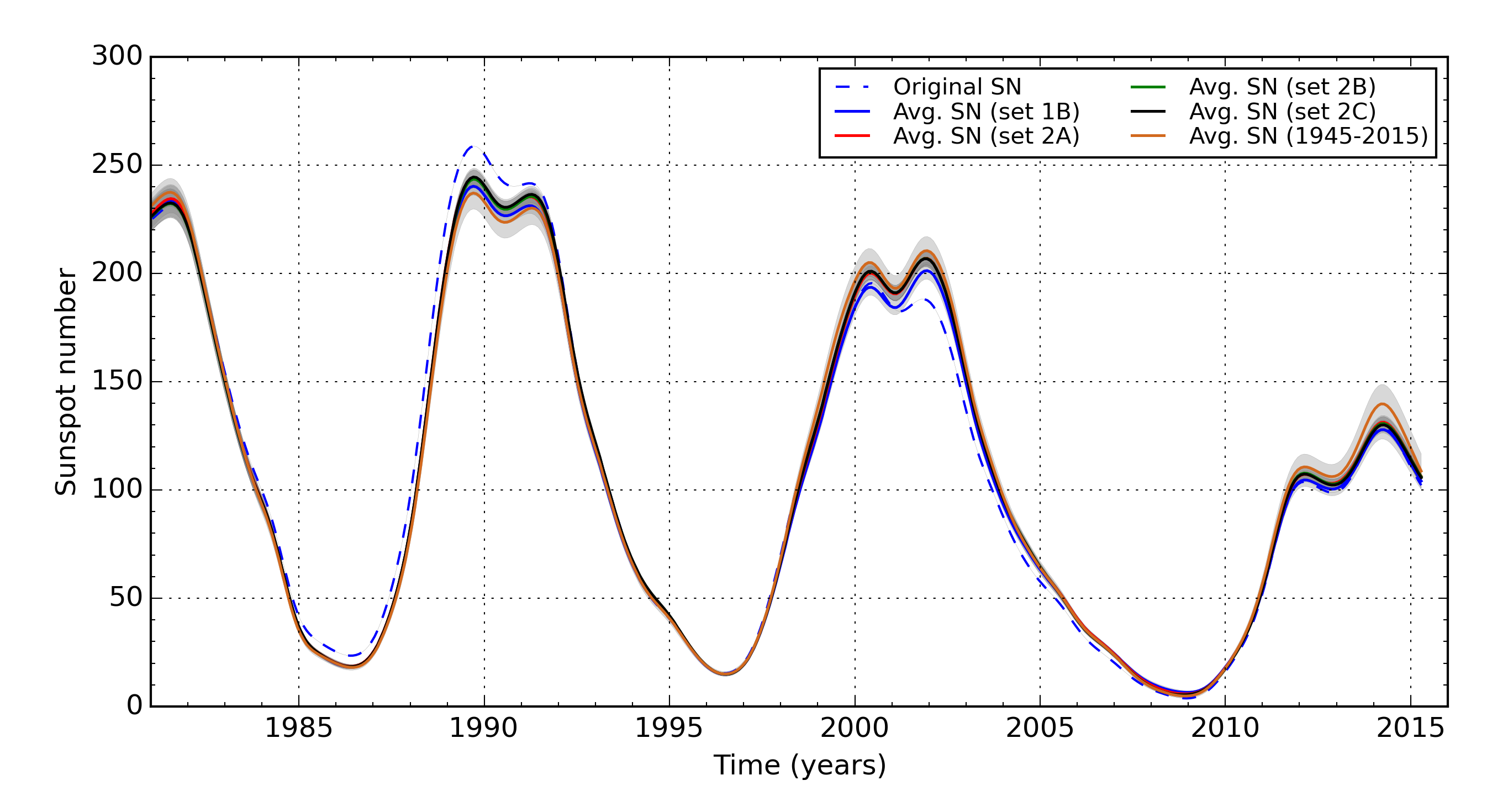}}
	\caption{Comparison of several average SN series based on different sets of stations, ranging from the entire set of available stations, including stations of only moderate quality (set1B, 2A; 72 stations) to a lower number of high quality stations (sets 2B, 2C; 42 stations). All curves closely agree, within the standard error (grey shading), thus showing that the reconstruction does not depend on the chosen station set. On the other hand, significant disagreements can be observed with the original SN series (blue dashed curve).} 
	\label{Fig-comparsn}
\end{figure}

As mentioned before, we applied the same calculation to several sets of station ranked in terms of quality (Fig. \ref{Fig-comparsn}). Although the largest set with the lowest quality threshold deviates a bit more (set 2A in figure \ref{Fig-comparsn}), the resulting average SN series all agree within the uncertainties. We thus conclude that the exact choice of stations does not lead to significant differences in the end result. We finally chose as standard station set, an intermediate selection containing 42 stations with a correlation coefficient $\rm r > 0.975$ (coefficient of determination $r^2 > 0.95$; Table \ref{tab_statselec}). 

\begin{table}
\caption{List of stations selected from the 280 stations in the WDC-SILSO database for the determination of the reference k correction over 1981--2015. The stations are listed by decreasing number of observed months. For each station, we give its k personal coefficient relative to the final reconstructed average SN, the error on k, the number of observed months (NM) and the coefficient of determination $r^2$ (quality of fit).}
\label{tab_statselec}
	
\begin{tabular}{llrrrr}     
\hline
ID & Station                      &  \multicolumn{1}{c}{k} & \multicolumn{1}{c}{$\sigma_k$} 
		& \multicolumn{1}{c}{NM} & \multicolumn{1}{c}{$r^2$}  \\
\hline
KZ & Kanzelh\"{o}he Obs., Austria         & 1.122 & 0.005 & 413 & 0.978  \\
MC & R.MacKenzie, Dover, UK               & 1.610 & 0.009 & 413 & 0.967  \\
SZ & M. Suzuki, Japan                     & 0.832 & 0.005 & 412 & 0.969  \\
UC & Uccle Obs., Brussels, Belgium        & 1.295 & 0.007 & 412 & 0.972  \\
FU & K.Fujimori, Nagano, Japan            & 1.179 & 0.006 & 410 & 0.970  \\
MO & E.Mochizuki, Urawa Saitama, Japan   & 1.127 & 0.006 & 409 & 0.966  \\
NY & Nijmegen, Netherlands                & 1.287 & 0.007 & 385 & 0.975  \\
DU & F.Dubois, Langemark, Belgium         & 1.224 & 0.006 & 365 & 0.977  \\
NZ & H.Barnes, Auckland, New Zealand      & 1.440 & 0.009 & 365 & 0.972  \\
KA & Kawagushi Obs., Japan                & 1.007 & 0.003 & 351 & 0.990  \\
PK & U.Pasternak, Berlin, Germany         & 1.547 & 0.011 & 347 & 0.963  \\
KS & Kislovodsk Obs., Russia              & 1.282 & 0.007 & 334 & 0.976  \\
CL & L.Claeys, Vedrin, Belgium            & 1.323 & 0.008 & 321 & 0.970  \\
AP & A.Philippe, Wittelsheim, France      & 1.429 & 0.010 & 313 & 0.959  \\
LS & Larissa Obs., Greece                 & 1.598 & 0.010 & 298 & 0.967  \\
SO & Sobota Obs., Slovakia                & 1.050 & 0.006 & 279 & 0.980  \\
CC & C.Courdurie, Marq-en-Bar{\oe}ul, France & 1.478 & 0.010 & 270 & 0.973  \\
BL & J.M.Bullon, Valencia, Spain          & 1.079 & 0.010 & 269 & 0.956  \\
JB & J.Bourgeois, Ciney, Belgium          & 1.120 & 0.009 & 266 & 0.963  \\
HP & Hvezdaren, Presov, Slovakia          & 1.142 & 0.007 & 257 & 0.974  \\
HV & Hvezdaren, Kysucke, Slovakia         & 1.085 & 0.008 & 251 & 0.971  \\
RA & Ramey, Puerto Rico                   & 1.104 & 0.007 & 239 & 0.970  \\
GL & G. Morales, Cochabamba, Bolivia      & 1.009 & 0.008 & 230 & 0.968  \\
AU & Coonabarabran Obs., Australia        & 1.627 & 0.012 & 221 & 0.966  \\
AM & Boulder, USA                         & 1.094 & 0.004 & 207 & 0.993  \\
BY & Beyazit Obs., Turkey                 & 1.259 & 0.009 & 205 & 0.975  \\
A2 & Athens Obs., Greece                  & 1.683 & 0.015 & 204 & 0.959  \\
VE & Ventura, Mosta, Malta                & 1.720 & 0.010 & 197 & 0.980  \\
YV & D.Yvergneaux, Renaix, Belgium        & 1.640 & 0.012 & 197 & 0.972  \\
KO & H.Koyama, Japan                      & 1.286 & 0.009 & 182 & 0.977  \\
AW & P.Ahnert Obs., Cottbus, Germany      & 1.084 & 0.006 & 179 & 0.984  \\
GA & G.Araujo, Spain                      & 1.035 & 0.007 & 176 & 0.977  \\
PA & Palehua, Hawai, USA                  & 1.177 & 0.009 & 161 & 0.980  \\
GS & G.-L.Schott, Germany                 & 1.728 & 0.015 & 158 & 0.972  \\
JP & Havana Sta., Cuba                    & 1.557 & 0.012 & 156 & 0.978  \\
MN & R. de Manzano, Italy                 & 1.102 & 0.011 & 144 & 0.968  \\
AA & NOAA/ISOON, USA                      & 1.485 & 0.016 & 131 & 0.956  \\
TY & T. Tanti, Malta                      & 1.469 & 0.014 & 123 & 0.970  \\
CO & Crimean Obs. Ukraine                 & 1.326 & 0.015 & 117 & 0.961  \\
TR & Trieste Obs., Italy                  & 1.819 & 0.019 & 108 & 0.968  \\
DM & G.Deman, Belgium                     & 1.132 & 0.013 & 107 & 0.963  \\
A3 & Elias, Athens, Greece                & 1.131 & 0.009 & 107 & 0.984  \\		
\hline
\end{tabular}
\end{table}

For the determination of the k ratio, we applied 6 different methods in order to check the consistency of the results, namely:
\begin{itemize}
	\item Four linear fits: $W_S$ versus $S_N$, the reverse $S_N$ versus $W_S$, both for a normal linear fit and a fit forcing the intercept through the origin (corresponding to a simple scaling factor)
	\item the factor leading to a null mean difference between the two series, used e.g. in Lockwood et al. (2014)
	\item the simple average ratio, sometimes used in past studies but which is oversensitive to fluctuation in low SN values and thus to imprecise ratios during solar cycle minima.
\end{itemize} 

\begin{figure}
	\centerline{\includegraphics[width=0.8\textwidth,clip=true,trim= 15 0 5 0,clip=true]{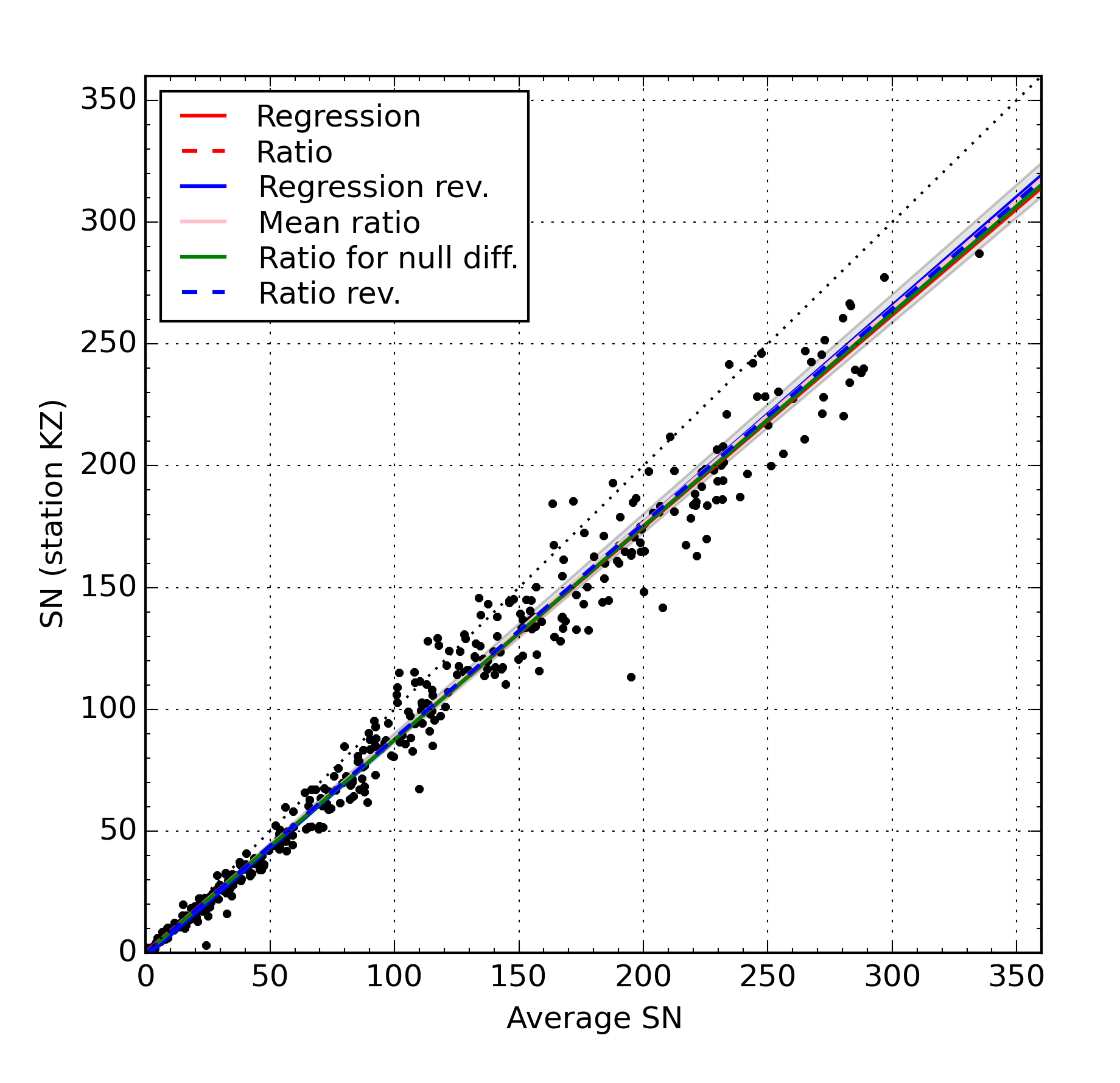}}
	\caption{Comparison of multiple determinations of the scaling ratio between the raw numbers from a single station and the average SN reference (Y/X and reverse X/Y linear fits, mean ratio, ratio giving a null total difference; see main text). For this moderately stable station (Kanzelh\"{o}he Observatory), giving a coefficient of determination $r^2 = 0.975$, the plot shows that all fits agree within the small uncertainty (gray shading= 1 standard deviation on the slope). Except for low-quality stations, the normalization of each station can be thus done very accurately by a standard linear fit.} 
	\label{Fig-qualfit}
\end{figure}

We find that for all good stations included in our data sets (coefficient of determination $r^2 > 0.90 $), all derived k factors match within the uncertainties (Fig. \ref{Fig-qualfit}).  Likewise, the linear fits pass through the origin, thus validating the hypothesis of a simple scaling factor between the series. Only the average ratio can deviate abnormally in a few cases, as can be expected from its oversensitivity to errors in low SN values. This crude estimator must thus be avoided. 

\begin{figure} 
	\centerline{\includegraphics[width=1.0\textwidth,clip=true,trim= 15 0 5 0,clip=true]{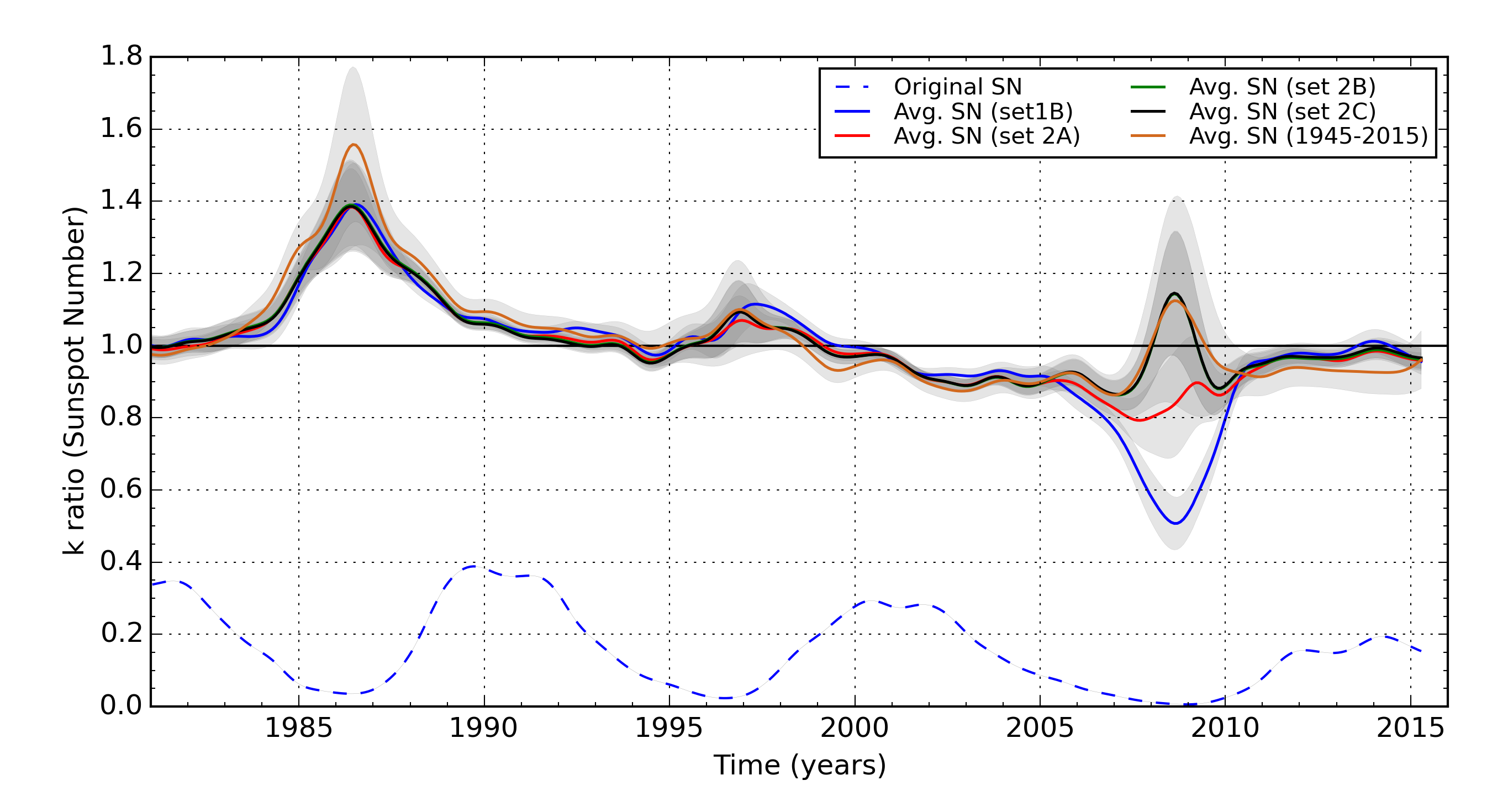}}
	\caption{k ratios between the average SN obtained from different sets of stations (cf. figure \ref{Fig-comparsn}) and the original SN series. The grey shading gives the standard error, which peaks at the time of solar cycle minima (the evolution of the cycles is given below by the dashed curve). We can see that the original sunspot number was first overestimated between 1983 and 1994 then underestimated after 2000 before returning close to the 1981 level in recent years. This confirms and refines the initial determination of the drift due to the Locarno pilot station diagnosed by Clette et al. (2014). The brown curve labeled (1945-2015) results from the use of the long-duration analysis described in section \ref{S-SN1945} and shows a very good agreement with this 1981-2015 analysis, which benefits from a larger sample of stations.}
	\label{Fig-compark}
\end{figure}

Having obtained this reconstruction of the SN, we can compute the ratio with the original SN, which gives the k correction factor shown in figure \ref{Fig-compark}. Comparing this new determination with the first version of the k ratio \citep[Fig. 52]{Clette_etal_2014}, we conclude that the two k series largely agree within the uncertainties. The largest differences occur during cycle minima, where the errors are largest but correspond only to minor differences in SN values.  All the features and the amplitude of the Locarno drift are thus confirmed (initial overestimate, decline and a plateau from 1990 to 1995, return to the 1981 scale after 2008 ). We only note that the new reconstruction indicates that the first upward deviation starts a bit later (1983) and more sharply, and that the overestimate over cycle 22 is slightly lowered. 

As the Locarno observers are applying the weighting according to spot size introduced by the Z\"{u}rich Observatory in their spot counts, we checked if the corresponding inflation in their Wolf numbers can account for some of the 1981-2015 drifts, as it produces a solar cycle variation in the ratio with standard Wolf numbers (all spots counted as 1). In order to check this, we applied the corresponding correction factor derived from simultaneous weighted and standard counts \citep[page 73, Fig. 44]{Clette_etal_2014}. 
\begin{equation}
 W= 1.123 + S_N / 1416    \label{EQ-weightfact}
\end{equation}

\begin{figure} 
	\centerline{\includegraphics[width=1.0\textwidth,clip=true,trim= 15 0 5 0,clip=true]{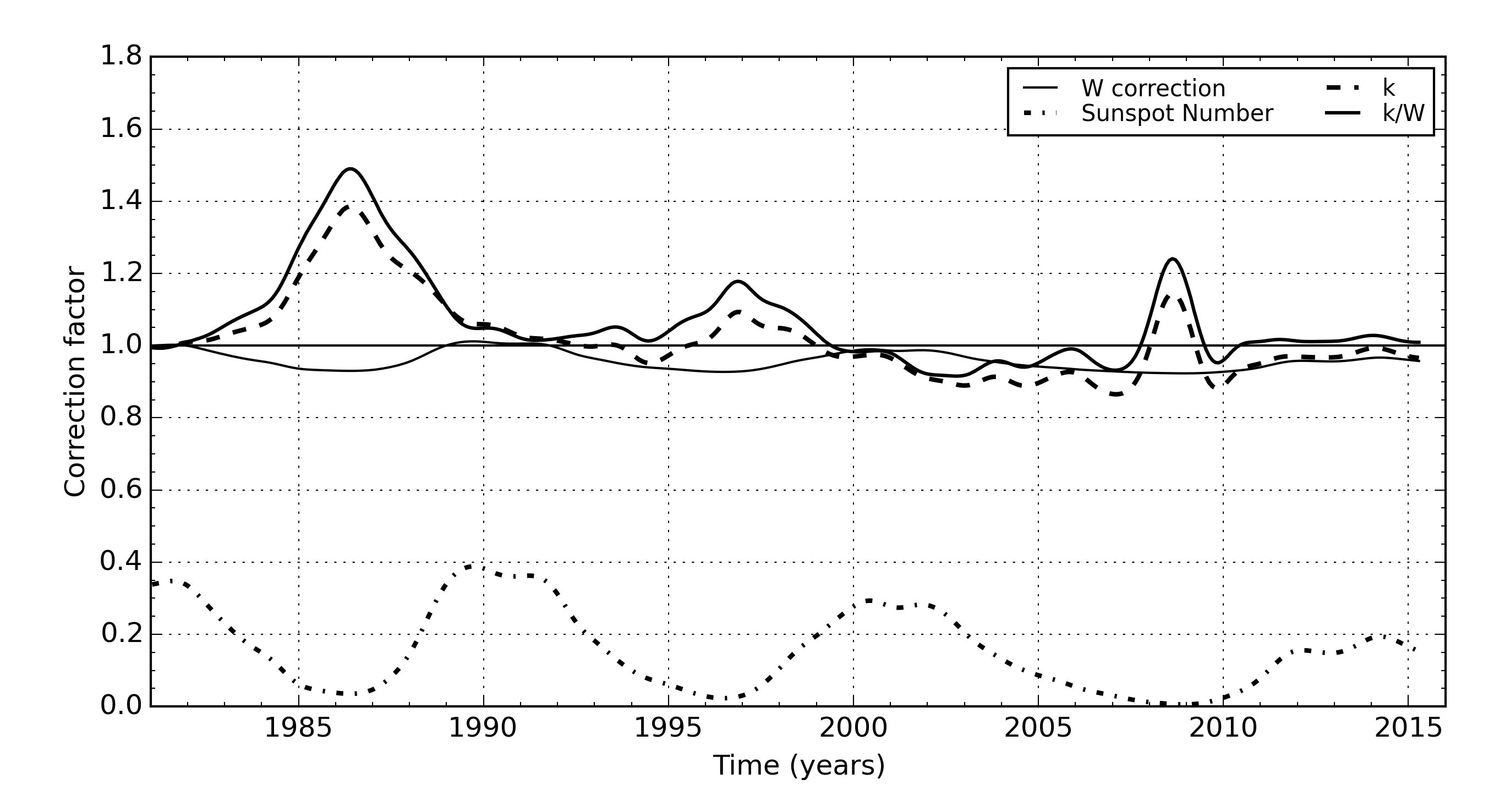}}
	\caption{Subtraction of the correction for the sunspot weighting according to spot size (thin solid line) from the average k correction determined here (dashed line). As the Locarno pilot station applied this spot weighting method, the result (solid line) shows the changes in the k ratio that are not associated with the weighting method. The weighting factor introduces a solar cycle modulation of about 10 \% (the SN is shown as a dash-dotted curve at the bottom) but it is smaller than the observed drifts, except after 2005, where most of the k correction seems to be explained by the weighting factor.} 
	\label{Fig-kweighting}
\end{figure}

Figure \ref{Fig-kweighting} shows the comparison of the original k factor and the corresponding factor after dividing k by the W correction. In principle, the latter reflects drifts that are not directly caused by the sunspot weighting. The overall variations are not changed. We only note that after elimination of the W factor, the overestimate becomes larger between 1983 and 1995 and the scale for recent years returns exactly at the level of 1981. So, the slightly lower counts currently provided by the Specola relative to 1981 seem to be only due to the effect of a lower sunspot weighting excess associated to the 50\% lower solar activity in cycle 24 than in cycle 21.

At this stage, the scale of the reconstructed SN is relative, as the initial common reference for step 1 on the iterative determination was the original Locarno series suffering from trends.
For the average k factor in figure \ref{Fig-compark}, we simply assumed that the initial scale was equal to the earlier Z\"{u}rich scale. Therefore, we adjusted the whole series by setting the average k factor to 1 over the first year (1981). As this adjustment rests only on a short time interval and some drift may already have taken place, we need to improve it and validate it. For this, we need to consider a longer time interval starting well before the 1981 transition.

\section{Reconstructed Sunspot Number 1945--2015} \label{S-SN1945}
\begin{figure} 
	\centerline{\includegraphics[width=1.0\textwidth,clip=true,trim= 15 0 5 0,clip=true]{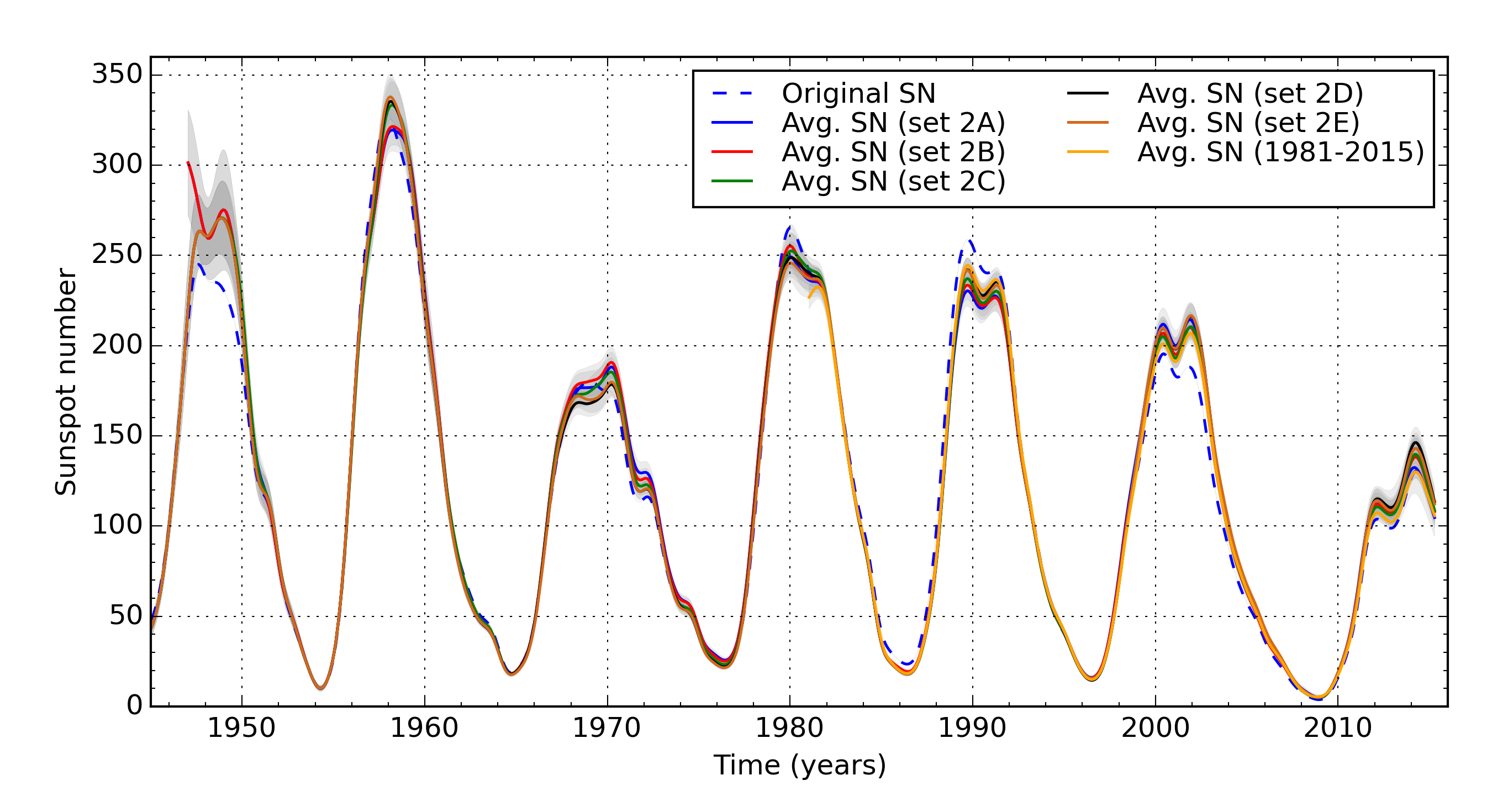}}
	\caption{Comparison of the reconstructed average SN series for different sets of stations, equivalent to figure \ref{Fig-comparsn} but for long-duration stations spanning the interval 1945-2015. Here the number of stations ranges from 8 (set 2A) to 35 (set2E). The 1981-2015 reconstruction (orange line) and the original sunspot number (blue dashed line) are included for comparison. Again here, all average SNs agree within the uncertainties, except before 1952, due to the low number of available data (1 or 2 stations).} 
	\label{Fig-compavgsn1945}
\end{figure}

We thus applied the same data selection and 2-step calculation to the long data set formed by the recovered long-duration stations described in section  \ref{S-Data}. In this case, we used the Kislovodsk Wolf number series as initial reference, given its long duration (1950-2015) and good stability. Again here, the reconstructed SN resulting from different station sets spanning different quality ranges lead to almost identical results, except for years before 1950, where the number of stations is rather low (down to 1 in some sets). This is reflected by larger uncertainties for those years in figure \ref{Fig-compavgsn1945}. This early part of this ``backbone'' is thus less reliable and should be ignored in comparisons.

\begin{figure} 
	\centerline{\includegraphics[width=1.0\textwidth,clip=true,trim= 15 0 5 0,clip=true]{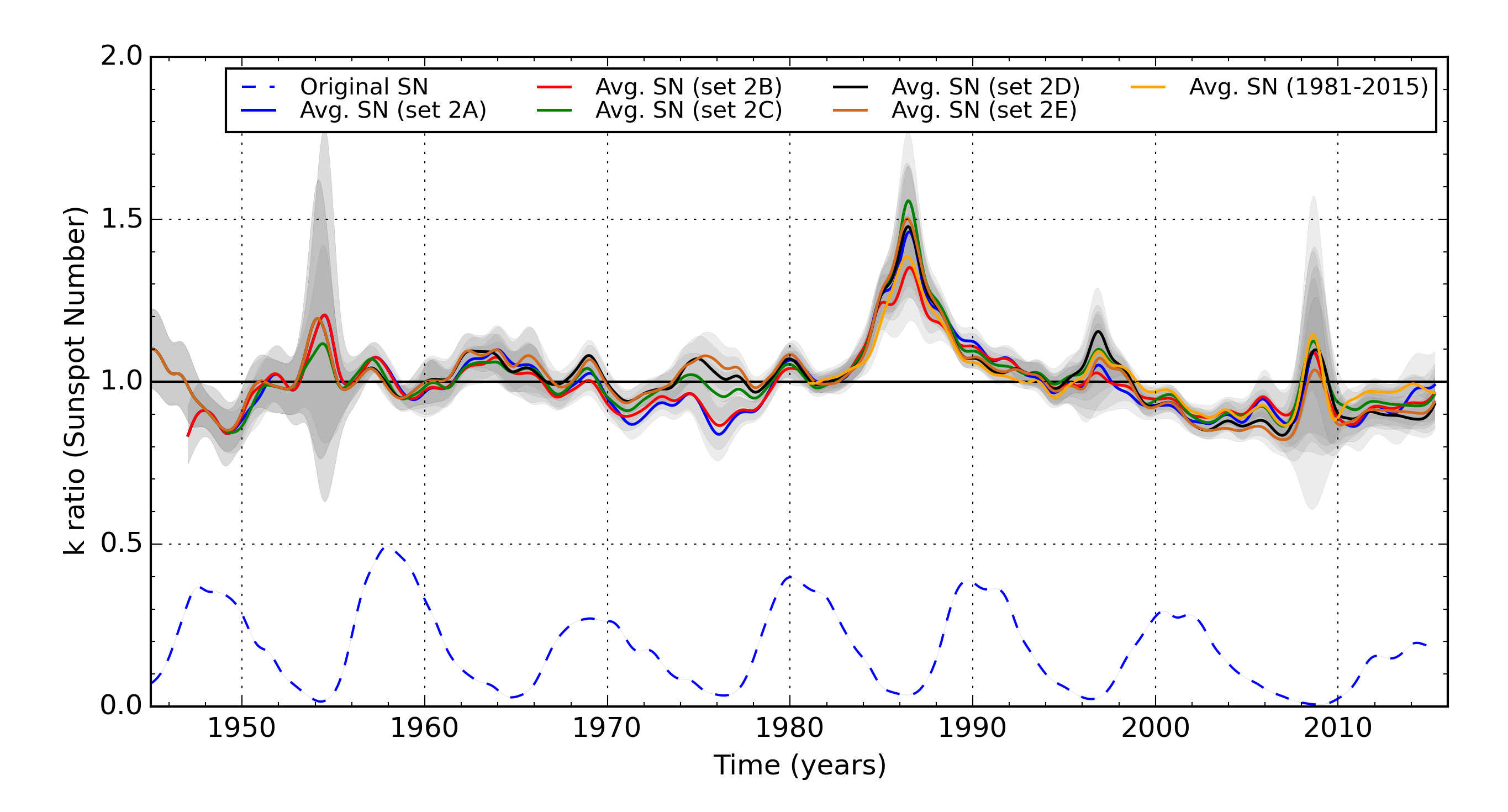}}
	\caption{Comparison of average k ratios between the reconstructed SNs for all data sets shown in figure \ref{Fig-compavgsn1945} and the original SN series. All k ratios closely agree within the standard error (gray shading), with larger uncertainties before 1955 and after 2008, due to the low number of available stations. The k variations found after 1983 are clearly visible while the k ratio is relatively stable and compatible with a unit ratio before 1983.} 
	\label{Fig-compavgk1945}
\end{figure}

Comparing the reconstructed SN with the original SN, we recognize the Locarno drift and again find that it starts only in 1983 (Fig. \ref{Fig-compavgk1945}). A comparison with the 1981-2015 reconstruction described in the previous section also shows a good agreement between the two series, except for a divergence in the very recent years where the long 1945-2015 data sets becomes less reliable, as mentioned above.

\begin{figure} 
	\centerline{\includegraphics[width=1.0\textwidth,clip=true,trim= 15 0 5 0,clip=true]{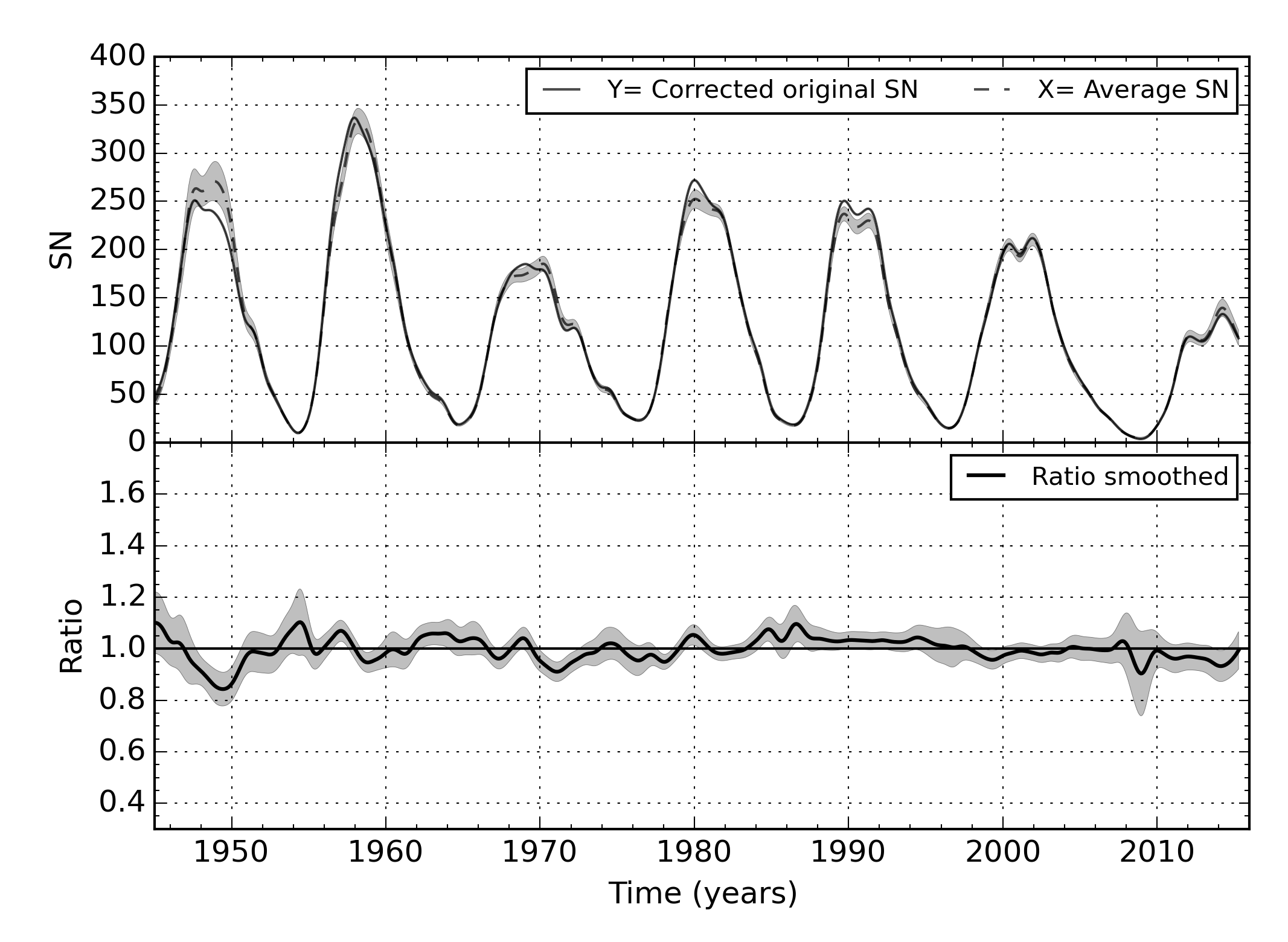}}
	\caption{Comparison between the average SN reconstructed over 1945-2015 (solid line) and the original SN series corrected by the k correction determined over the 1981-2015 interval (dashed line in top panel). The resulting ratio (lower panel) is now constant within the standard error (grey shading). Moreover, there is no significant change between the average ratio before and after the 1981 Z\"{u}rich--Locarno transition, thus indicating that the absolute scale of the 1981--2015 correction matches the scale of the Z\"{u}rich period before 1981.} \label{Fig-sncorr1981}
\end{figure}

We note that the k ratio also displays fluctuations in the pre-1980 Z\"urich era, but only temporary and of lower amplitude than after 1983. No significant long-duration trend is affecting this part. If we now apply the 1981-2015 k correction factor to the second part of the new 1944-2015 SN (Fig. \ref{Fig-sncorr1981}), the post-1981 trends vanish and the average ratio becomes constant for the 1981-2015 interval, confirming that the defects in the original SN series are now largely eliminated.

As this now-uniform series straddles the 1981 transition, we can check that the preliminary scale adopted for the 1981-2015 correction (see previous section) is validated by the long-duration series. For this, we derive the ratio between the average k computed respectively over the entire intervals 1945-1980 and 1981-2015. Taking the ratio between those to mean k values, we find that the corrected 1981--2015 SN is barely higher than the 1945-1980 SN, by a factor $1.005 \pm 0.006$. The difference with unity is thus not significant and we can adopt this correction with an overall uncertainty of 1 \%.

At this stage, we matched the scale of the Brussels-Locarno SN after 1981 to the scale of the Z\"{u}rich SN prior to the 1980 transition. Still, we must remember that the latter is itself affected by a scaling bias resulting from the adoption of weighted spot counts by Waldmeier in 1947 and that this factor might be variable according to the level of solar activity, as suggested by dual counts (equation \ref{EQ-weightfact}). So, just like for the original SN, the present reconstruction still propagates the scale of the inflated 1947--1980 SN to the present values. The final correction of the sunspot weighting bias will be treated in a subsequent paper \citep[Paper II]{Clette_etal_2015} and through this new 1945-2015 ``backbone'' SN will allow to standardize the scale over the entire SN series.

\section{Reconstructed Group Number} \label{S-RecGN}
In order to obtain an equivalent GN series, we applied the same analysis but using raw group counts instead of raw Wolf numbers from exactly the same sets of stations. For both analyses, we used the Specola Locarno station as reference, as we assumed that the inhomogeneities in the group counts are much weaker than in Wolf numbers, as the detection of small spots and the weighting according to spot size mainly influence the spot counts.

Like for the SN, we ranked the stations according to the degree of correlation with the average series and the residual standard deviation. With only a few exceptions, this led to the same ranking as for the SN. Therefore, we adopted the same sets of stations as for the SN analysis, which ensure a total identity of the data sets. Overall, Group Numbers showed a higher degree of correlation, as can be expected when excluding the counts of individual spots which differ more from observer to observer. 

\begin{figure} 
	\centerline{\includegraphics[width=1.0\textwidth,clip=true,trim= 15 0 5 0,clip=true]{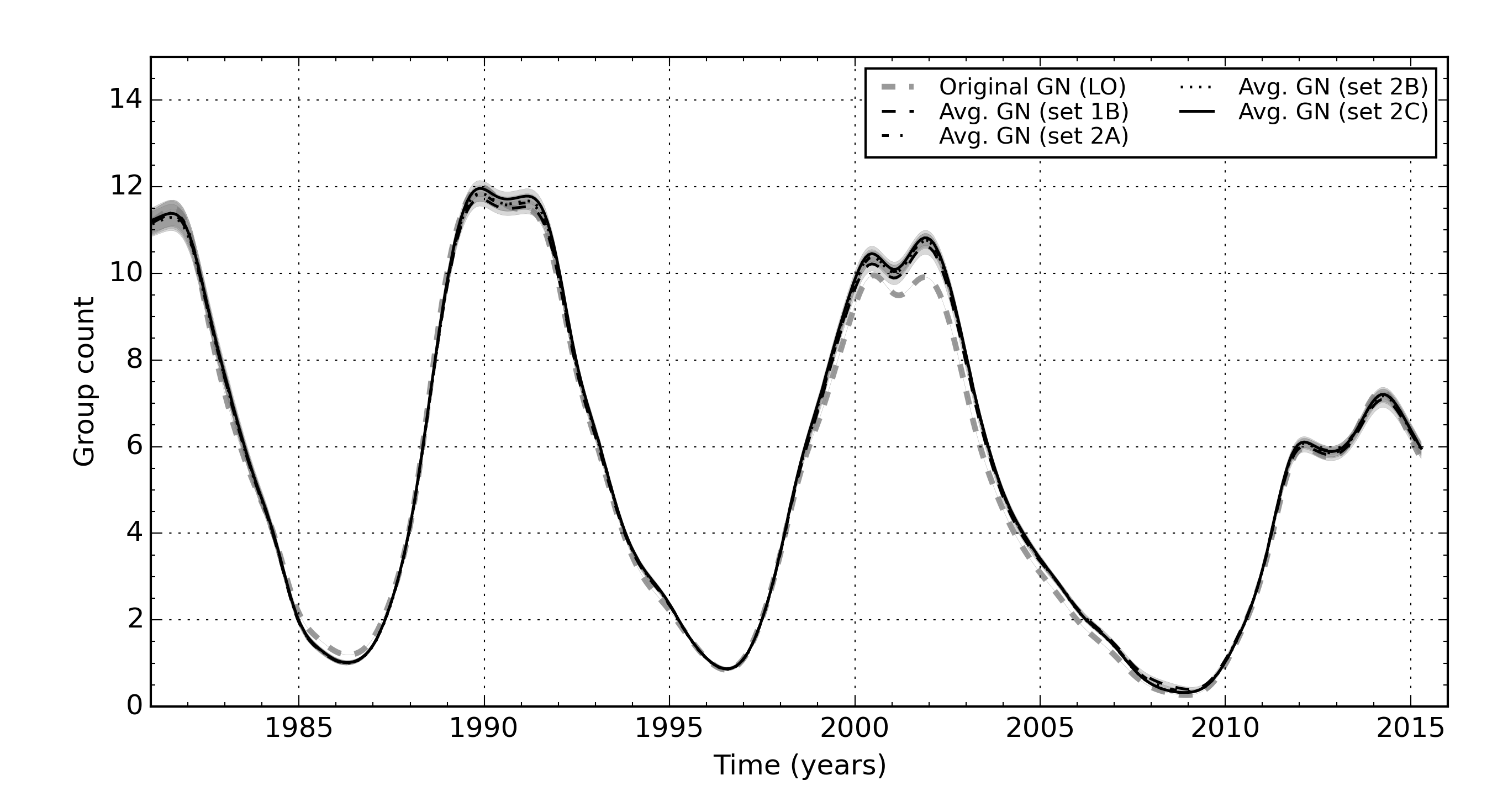}}
	\caption{Reconstructed average GN series over the 1981-2015 interval for the same sets of stations as for the SN (cf. figure \ref{Fig-comparsn}). The agreement between the series is even closer than for the SN. The gray dashed line shows the base GN series from the Locarno reference station. It shows a significant underestimate between 1996 and 2010.} 
	\label{Fig-compavgGN}
\end{figure}

\begin{figure} 
	\centerline{\includegraphics[width=1.0\textwidth,clip=true,trim= 15 0 5 0,clip=true]{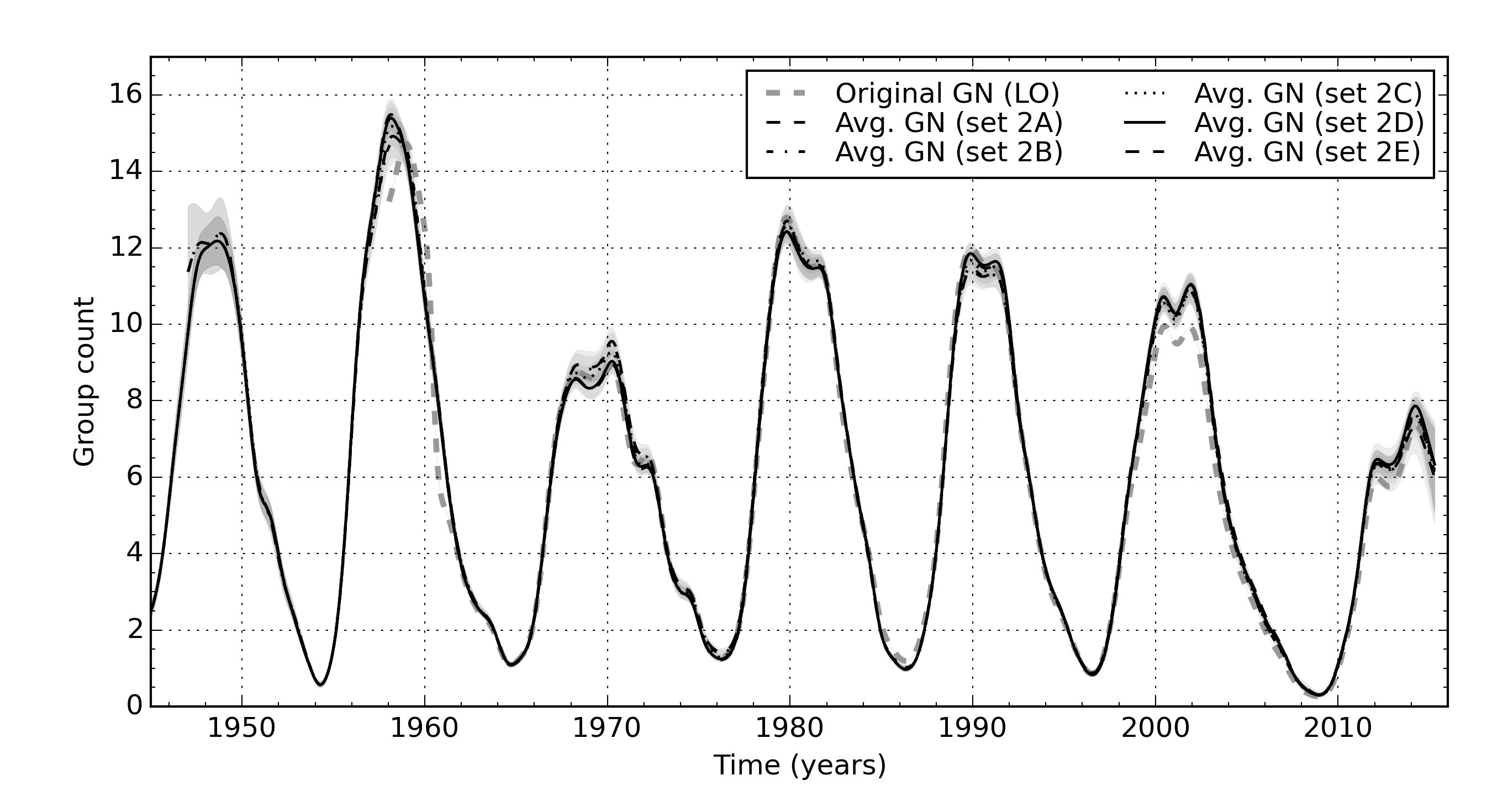}}
	\caption{Reconstructed average GN series over the long 1945-2015 interval for the same sets of stations as for the SN (cf. figure \ref{Fig-compavgsn1945}). Like for the SN, larger errors before 1950 and after 2010 are due to the limited number of stations. The gray dashed line shows the base GN series from the Locarno reference station. Significant disagreements can be see in the early years over 1958--1963 (learning period?) and again over 1996--2010.} 
	\label{Fig-compavgGN1945}
\end{figure}

\begin{figure} 
	\centerline{\includegraphics[width=1.0\textwidth,clip=true,trim= 15 0 5 0,clip=true]{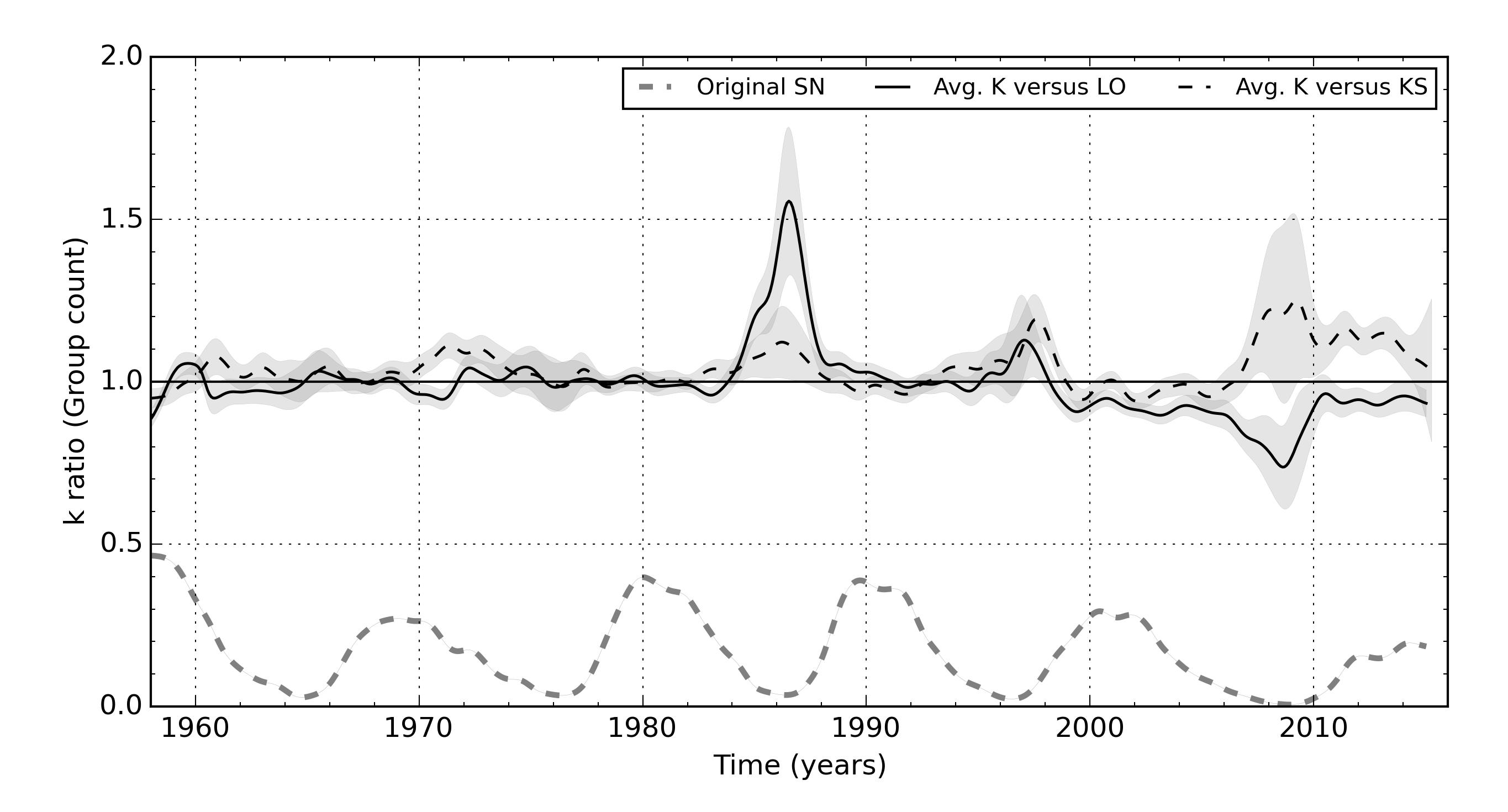}}
	\caption{k ratio between the Locarno group counts and the reconstructed GN (solid line) and as a comparison the equivalent ratio  for the Kislovodsk station (dashed line). The original SN is shown at the bottom (thick grey dashed line) to indicate the solar cycle timebase. The Kislovodsk GNs are very stable until 2007, when they jump to a higher scale, indicating a change of counting technique. On the other hand, the LO GNs start to deviate in 1983 and follow an evolution similar to the corresponding SNs, although the drifts have a smaller amplitude for the GN.} 
	\label{Fig-compk_loks}
\end{figure}

Here again, the reconstructed series match closely, regardless of the set of stations included in the average, as can be seen in figures \ref{Fig-compavgGN} and  \ref{Fig-compavgGN1945}. When checking the stability of the Locarno reference, we still find the same drifts after 1980, though with a lower amplitude (Fig. \ref{Fig-compk_loks}). As a comparison, we can see that Kislovodsk does not show such marked trends and was thus more stable.

\begin{figure} 
	\centerline{\includegraphics[width=1.0\textwidth,clip=true,trim= 15 0 5 0,clip=true]{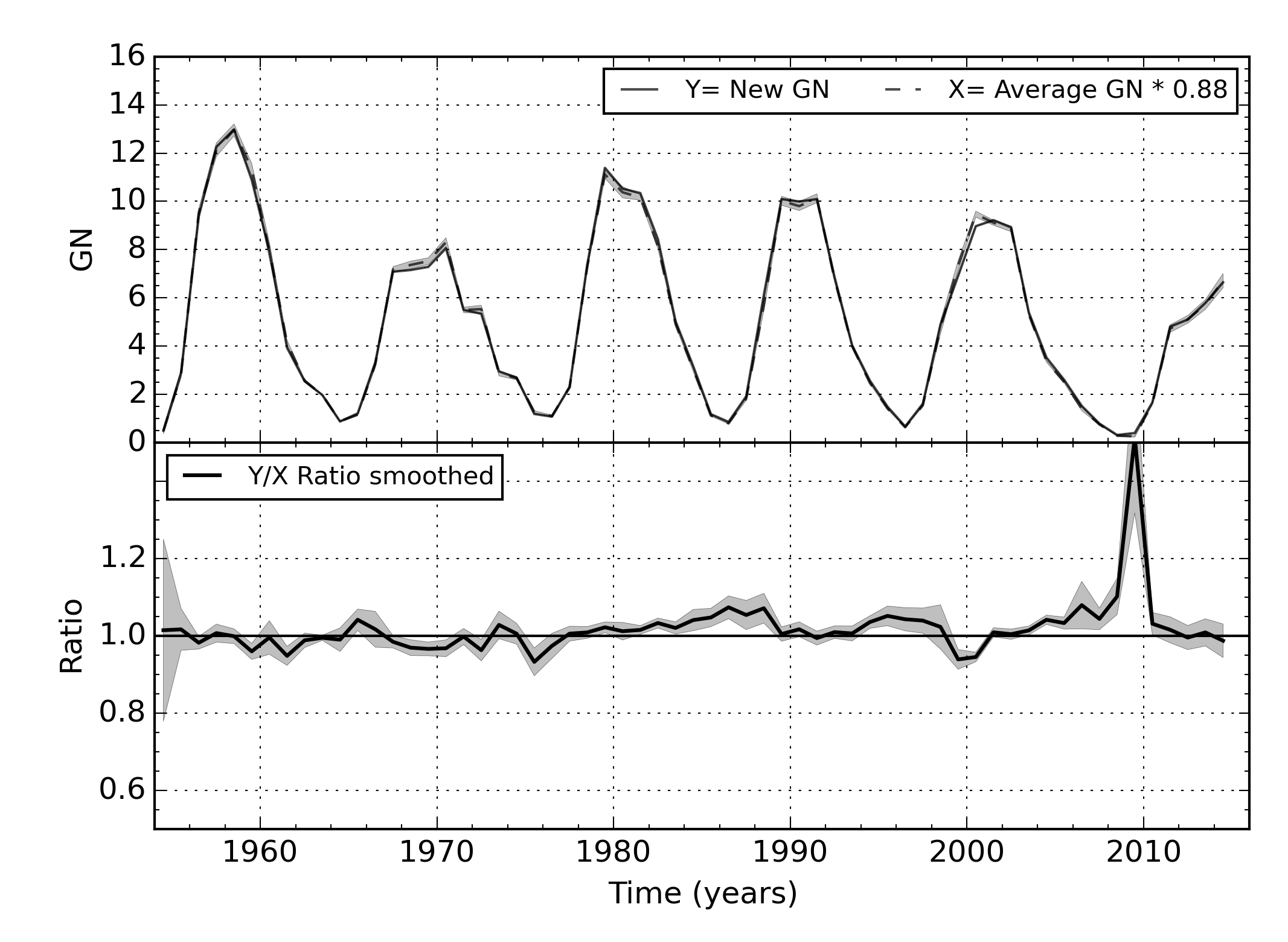}}
	\caption{Comparison between our new reconstructed GN (dashed line) and the new recalibrated ``backbone'' GN (solid line) \citep{Svalgaard-Schatten_2015, Clette_etal_2014}. Both series closely agree within the uncertainties, leading to constant ratio over the entire 1945-2015 interval (lower panel). A scale factor of 0.88 was applied to the reconstructed GN (normalized to the Locarno group counts) to bring it to the scale of the new GN (normalized to Wolfer's numbers in the early $20^{th}$ century). The peak in 2009 is an artifact due to the large number of spotless days in the deep minimum (very low group counts). } 
	\label{Fig-compgn-gnbb}
\end{figure}

We can now compare this new reconstruction of the GN with the new reference GN series published by \citet[this issue]{Svalgaard-Schatten_2015}, covering the entire interval 1610--2015. We find that although both series use slightly different data sets and are derived by different calculations, the agreement between the two series is very high, with an overall degree of correlation of 0.998 (Fig. \ref{Fig-compgn-gnbb}). The ratio is uniform over the entire 1950-2015 interval (ignoring the inaccurate 1945-1950 section in our reconstruction). This gives a strong mutual validation between the two series. The average ratio between the two series indicates that our GN must be lowered by the factor $0.88 \pm 0.01$ to be brought to the Wolfer scale adopted as reference for the new official GN. This comes very close to the $0.91 \pm 0.01$ factor used by \citet{Svalgaard-Schatten_2015} for normalizing its ``Locarno backbone''. 
	
\begin{figure} 
	\centerline{\includegraphics[width=1.0\textwidth,clip=true,trim= 15 0 5 0,clip=true]{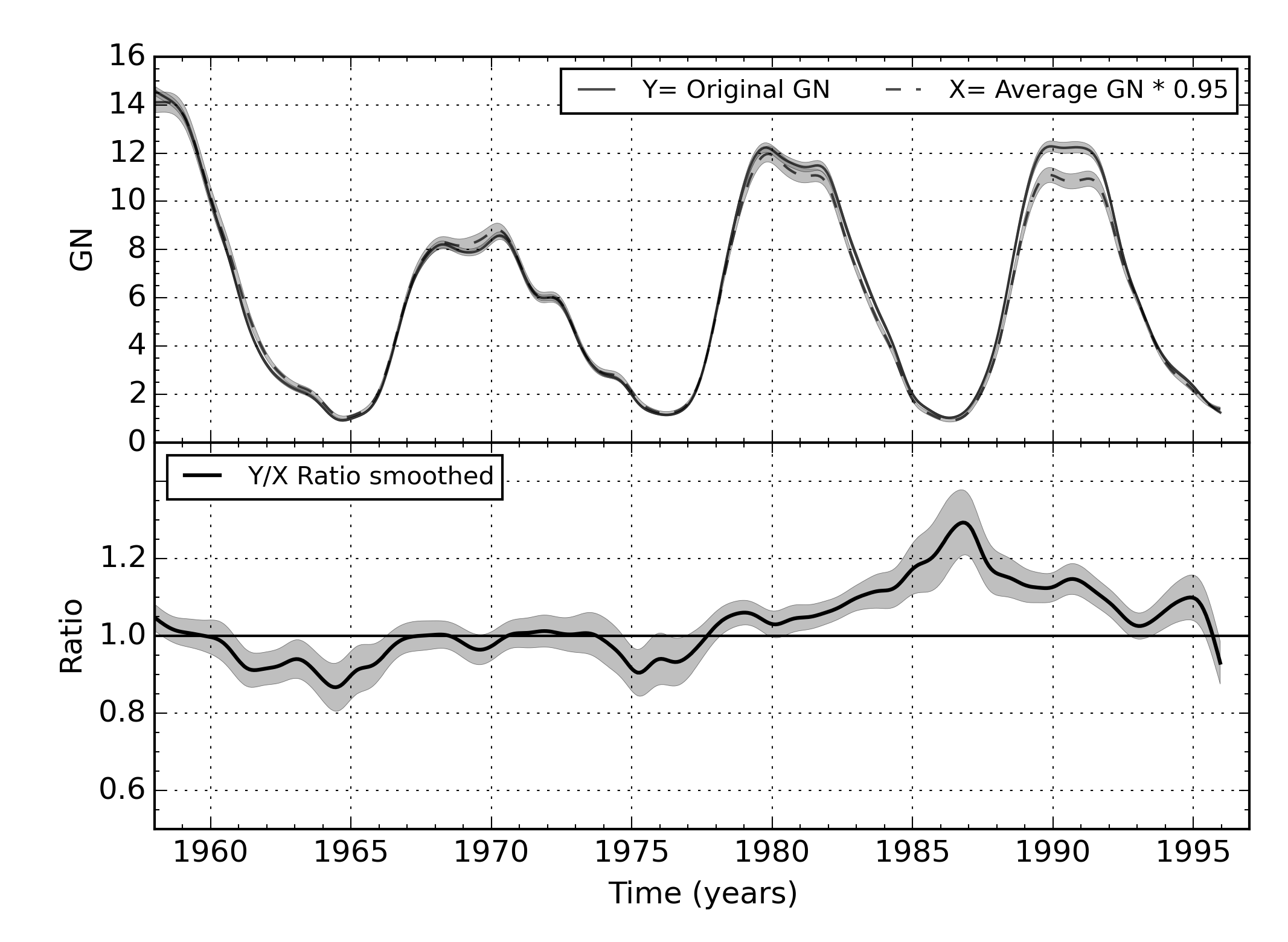}}
	\caption{Comparison of our reconstructed average GN (dashed line) with the original GN (solid line) \citep{Hoyt-Schatten_1998a, Hoyt-Schatten_1998b}. Here, we divided the original GN by the conventional 12.08 factor to reduce it back to the scale of regular group counts. We also multiplied the new GN by a factor 0.95 to match the scales before 1975. In the lower panel, the ratio starts to drift progressively after 1977, the original GN being then overestimated by up to 20 \%. } 
	\label{Fig-compgn-gnhs}
\end{figure}

We can now also compare in detail this new GN with the original GN produced by Hoyt and Schatten (1998). Recent analyses \citep{Lockwood_etal_2014, Clette_etal_2014} indicated that the scale of the original GN changes upward by 10 \% in 1976, when the photographic data from the Royal Greenwich Observatory come to an end and are succeeded by visual counts from the USAF/SOON stations in the original Hoyt and Schatten database \citep{Hoyt-Schatten_1998a, Hoyt-Schatten_1998b}. As can be seen in our new analysis (Fig. \ref{Fig-compgn-gnhs}), there is actually no sharp transition around 1976 but instead, the scale starts rising progressively to reach an excess of 12 \% in 1986, relative to the average scale before 1976. After this maximum, the scale is declining again until the end of the series in 1995. Therefore, the scale is not constant over those final years of the original GN and this can be explained by the transition from the Greenwich photographic catalog to the USAF/SOON visual network and by the decreasing number of included stations, down even to a single station in the final years. Consequently, the average scale over the interval 1977--1995 is only $8 \% \pm 0.5\%$ above the earlier scale for the $20^{th}$ century. This variable bias thus affects the scale of the whole original GN series, as Hoyt and Schatten used the average ratio between their GN and the original SN over the entire $20^{th}$ century.

\section{Group Number versus Sunspot Number: a constant non-linear relation} \label{S-SN-GNrel}

Now that we have both a GN and SN series built from exactly the same underlying data, we have for the first time the ability to study the mutual relation between those two indices, without any bias. Indeed, previous studies found systematic time variations in the GN/SN ratio \citep{Tlatov_2013, Clette_etal_2014}. Given the different information contained in those two numbers, this may reflect actual changes in solar properties over years and centuries. For instance, several studies show a decline in the SN/GN ratio since cycle 19, which seems to amplify after the maximum of cycle 23 around 2000 \citep{Svalgaard-Hudson_2010,Clette-Lefevre_2012,Svalgaard_2015b}. This suggests a corresponding decline in the average number of spots per groups. Given the unexpected low level of the current solar cycle, we may thus wonder if the acceleration in the decline of the SN/GN is a unique signature heralding a transition of the solar cycle towards a new regime, possibly an approaching Grand Minimum. However, past GN--SN comparisons used original GN and SN data based on different non-overlapping data sets and different stations. Therefore, variations due to discrepancies in the data may be misinterpreted as true solar variations and anyway, true changes in the SN--GN relations are obscured by non-matching inhomogeneities in both data series.

\begin{figure} 
	\centerline{\includegraphics[width=1.0\textwidth,clip=true,trim= 15 0 5 0,clip=true]{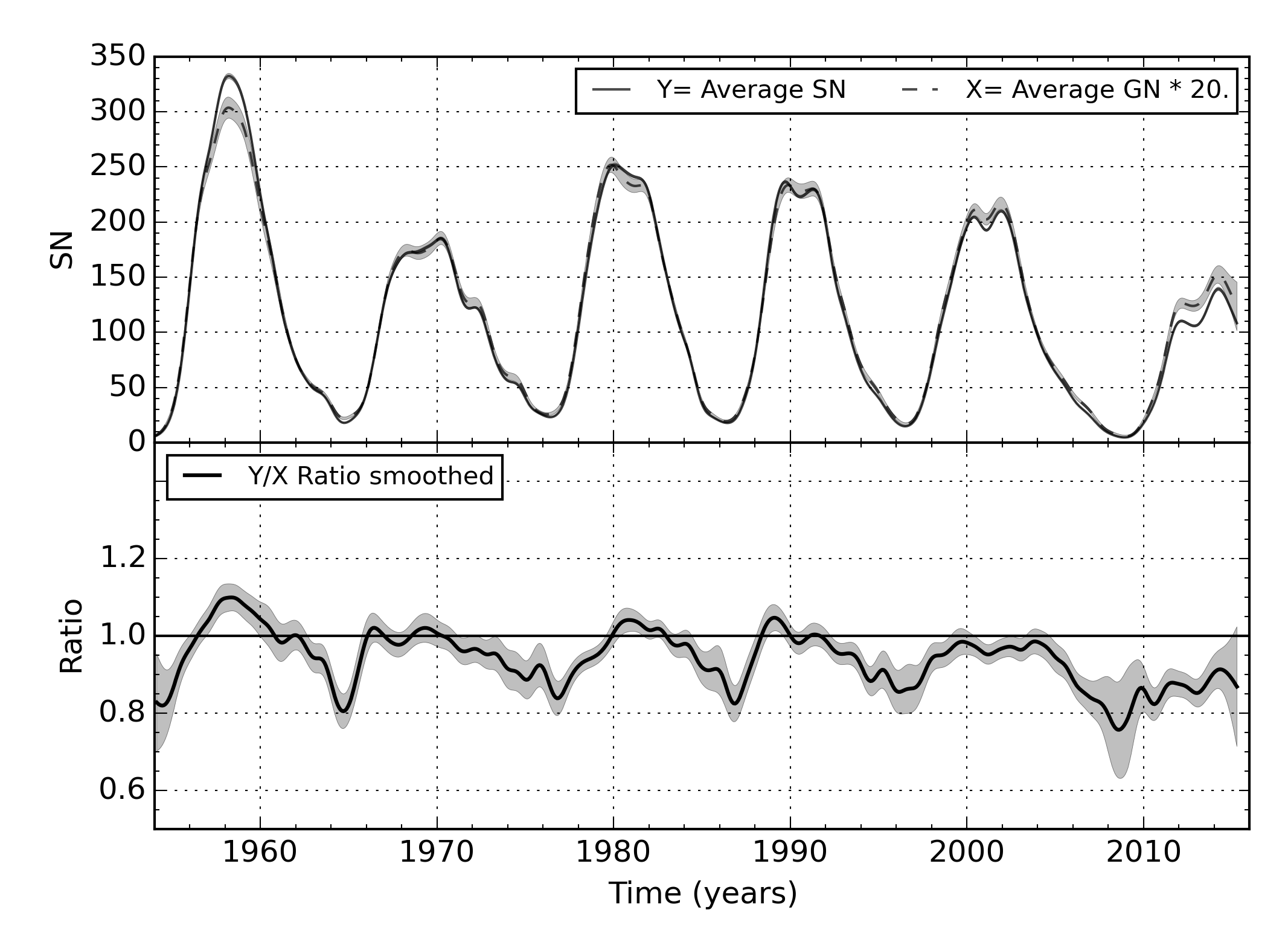}}
	\caption{Comparison between the reconstructed SN (solid line) and GN (dashed line), based on identical data sets over the 1945-2015 interval. To allow a direct comparison, the GN series was multiplied by a factor 20 to bring it to the same scale as the SN. Although both series show a good overall agreement, the ratio plotted in the lower panel shows a pronounced solar cycle variation as well as a downward trend from cycle 19 to 24, suggesting a variable relation between the two sunspot indices.} 
	\label{Fig-compsngn}
\end{figure}

\begin{figure} 
	\centerline{\includegraphics[width=0.8\textwidth,clip=true,trim= 15 0 5 0,clip=true]{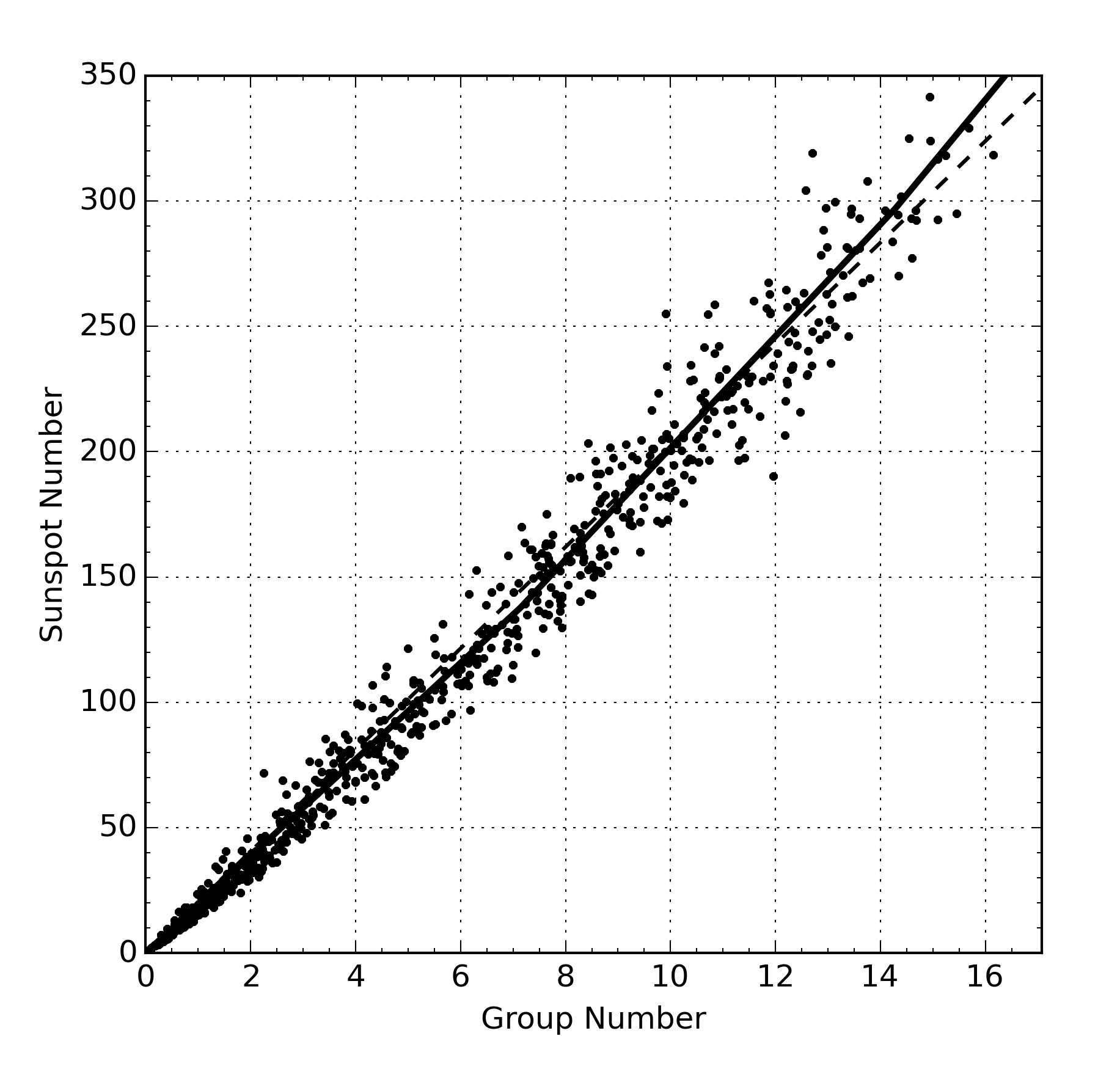}}
	\caption{Scatterplot of the monthly mean SN and GN. The SN does not follow a linear relation with the GN, with proportionally more sunspots per group as the level of activity increases. As a guide for simple proportionality, the dashed line gives the linear fit. The best fit is given by the second order polynomial in equation \ref{EQ-Poly2} (solid curve).} 
	\label{Fig-polfit}
\end{figure}

Figure \ref{Fig-compsngn} shows the comparison between our matching long SN and GN series over 1945-2015. We indeed observe an overall downward trend of the SN relative to the GN from 1960 to 2008, with the ratio dropping even lower for cycle 24. At the same time, we note that the trend matches the amplitude variations of the SN itself, i.e. the level of solar activity. This is confirmed over shorter timescales, as the ratio includes a prominent solar cycle modulation, indicating that the ratio is systematically higher for higher levels of activity. As the same behavior seems to be present over the entire time interval, we thus least-square fitted polynomials between the SN and the GN. We tried polynomials from degree 2 to 4, and found that degree 2 gave the best fit and highest degree of significance for the determined coefficients. The fitted degree 2 polynomial is shown in figure \ref{Fig-polfit}. Its expression is: 
\begin{equation}
 S_N = 17.8\,(\pm 0.4) \times G_N + 0.21\,(\pm 0.03) \times G_N^{2}  \label{EQ-Poly2}
\end{equation}
The positive second-degree coefficient indicates that the number of spots increases faster than the GN, with this trend amplifying as the activity level increases.
 
\begin{figure} 
	\centerline{\includegraphics[width=1.0\textwidth,clip=true,trim= 15 0 5 0,clip=true]{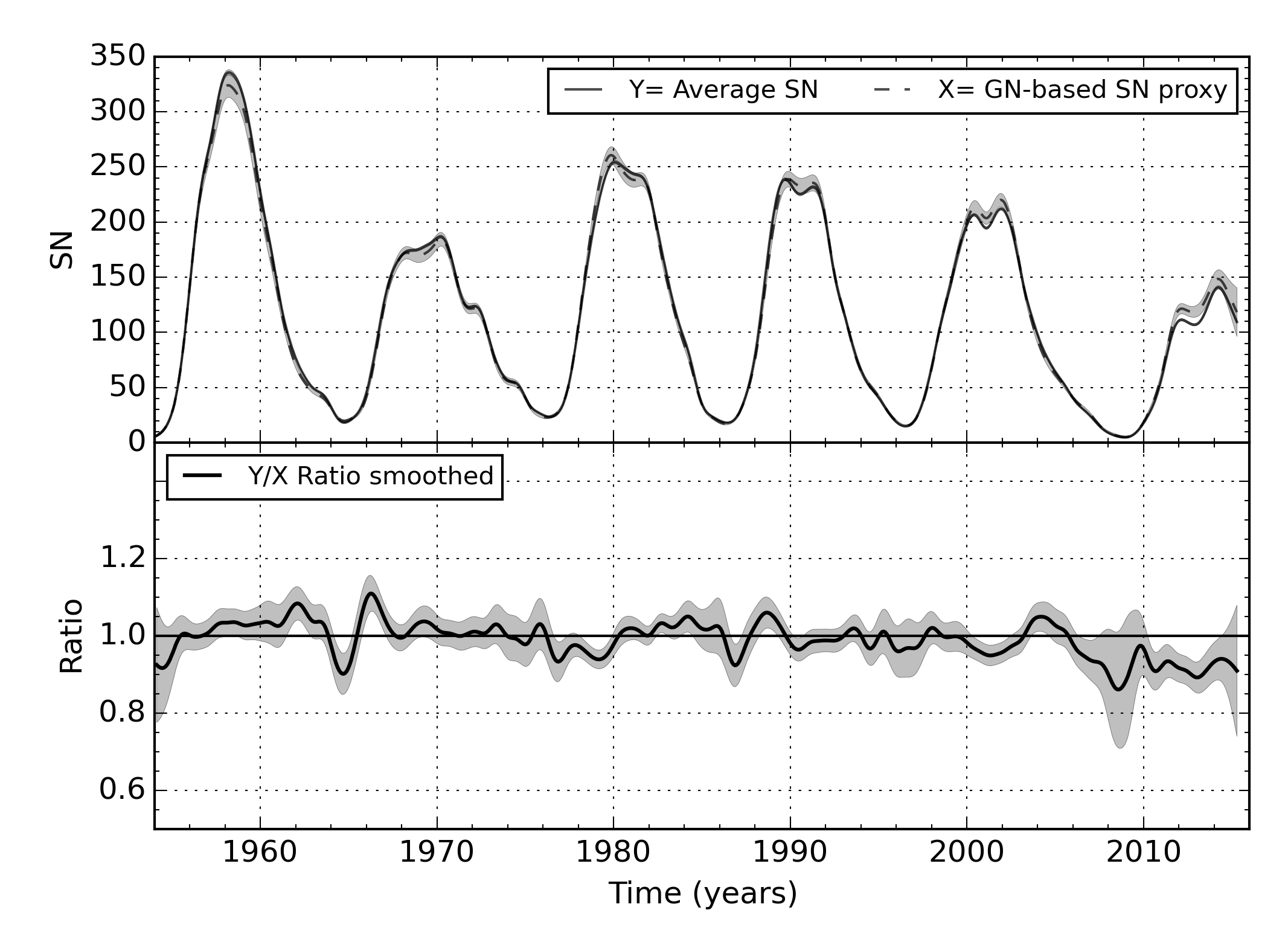}}
	\caption{Comparison between the average SN (solid line) and GN-based proxy obtained by equation \ref{EQ-Poly2} (dashed line). Compared to the base series shown in figure \ref{Fig-compsngn}, the ratio (lower panel) does not show a solar cycle variability anymore and there is no significant trend from cycles 19 to 23, in spite of the wide range of cycle amplitudes. Only cycle 24 differs from the other ones, with a ratio about 8\% lower, corresponding to a lower average number of spots per group.} 
	\label{Fig-compSNproxy}
\end{figure}

Having this non-linear conversion function, we can now build mutual proxies, either a GN proxy of SN or the reverse. The comparison between the new GN-based proxy and the corresponding SN series is shown in figure \ref{Fig-compSNproxy}. We can immediately see that the solar cycle modulation has largely vanished from the ratio. Moreover, the ratio is essentially constant over the entire time interval. We thus have the confirmation that this single non-linear relation accounts for most of the variations in the SN/GN ratio. Therefore, the apparently irregular time variations in the ratio between the SN and the GN result from a constant property in the sunspot population. In particular, the progressive declining trend between 1960 and 2008 does not imply any peculiar behavior of solar activity over that period.

However, a significant downward jump occurs in 2008, with the ratio then remaining stable at a value 0.925 ($\pm 0.008$) times lower than the preceding period over cycle 24. This jump is significant and indicates a true change in sunspot statistics in cycle 24. However, contrary to former results, this transition occurs only recently over the cycle 23-24 minimum and not progressively since cycle 19.

We still need to check the validity of this constant SN--GN relation over longer time interval, for past epochs in the $20^{th}$ century and even earlier. This will be verified in a subsequent paper (Paper II). However, as for past epochs, it is more difficult to work with identical data sets for the sunspot and group numbers, we can expect that the relation will be less marked, as random variations of the same magnitude as the non-linear correction will be superimposed. Therefore, the coherent 65-year long data set used in the present analysis delivers the most accurate determination of this relation.

\section{Comparisons with parallel standard reference series} \label{S-Indices}
Next to the GN, the two other solar times series to which the SN has always been first compared are $\rm F_{10.7}$, the 10.7 cm background radio flux, and $\rm R_a$, the American Sunspot Number maintained by the Solar Section of the American Association of Variable Stars Observers (AAVSO). Both series showed significant disagreements with the original SN series, leaving a doubt as to whether one of those series was flawed or the divergences revealed an intrinsic solar effect.

\subsection{Inhomogeneities in $\rm F_{10.7}$}  \label{SS-F107}
Several recent studies have shown that $\rm F_{10.7}$ and SN were in fair agreement until 1995 but then started to diverge, with the SN falling significantly below $\rm F_{10.7}$ after 2000 \citep{Svalgaard-Hudson_2010, Lukianova-Mursula_2011,Tapping-Valdes_2011, Clette-Lefevre_2012}. \citet{Clette_etal_2014} already showed that by applying the correction for the Locarno drifts, which precisely cause an underestimate of the SN after 2000, the difference between the two series was cut by about half, but that a significant disagreement was still present for that interval.

\begin{figure} 
	\centerline{\includegraphics[width=1.0\textwidth,clip=true,trim= 15 0 5 0,clip=true]{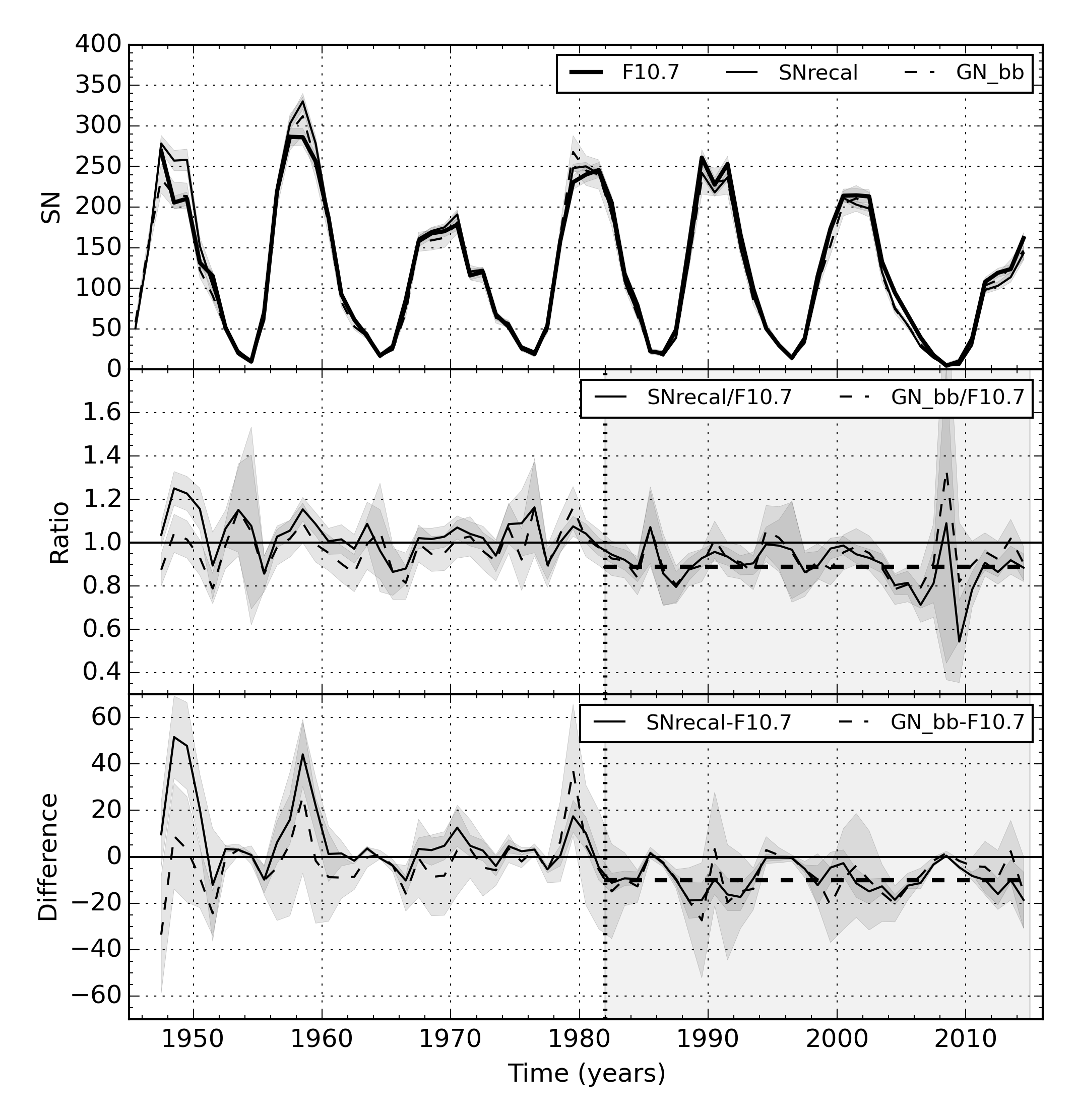}}
	\caption{Comparison of our reconstructed SN (solid line) and of the new GN series (dashed line) with the $\rm F_{10.7}$ radio flux (thick solid line; SN proxy by formula \ref{EQ-F107proxy}): the top panel shows the three series, the middle panel shows the two ratios and the bottom panel the differences. Standard errors are given by the grey shading. Except for stronger variations before 1955 attributable to a lower accuracy of our SN reconstruction, the ratios and differences are very similar for the SN and GN. After some known drifts in the early calibration of $\rm F_{10.7}$ until 1962, the ratios are essentially constant over the entire 68-year duration of the measurements. However, a definite change of scale occurs around 1983, with the $\rm F_{10.7}$ flux being systematically higher by 12 \%  after 1983 (shaded interval, and dashed horizontal line), leading to a drop of the ratio versus both the SN and GN compared to the early part.}. \label{Fig-compF107}
\end{figure}

Now, using the present reconstruction, either of the SN or of the GN (here, the  ``backbone'' reconstruction by \citet{Svalgaard-Schatten_2015}), we show the new comparison in figure \ref{Fig-compF107}, where we used the following $F_{10.7}$-based proxy formula, similar to formula R2 in \citet{Johnson_2011}:
\begin{equation}
   S_N(F_{10.7})= 4.677 \times (F_{10.7} - 68.0)^{0.81}   \label{EQ-F107proxy}
\end{equation}

The main feature that comes out of both comparisons is a change of scale happening between 1979 and 1983, with $\rm F_{10.7}$ flux about 12 \% ($\pm 1 \%$) higher in the second part of the series. In the first part, the flux is also slightly too low in the first years (1947- 1953), a calibration problem of the early data before a site change in 1962 \citep{Tapping-Morton_2013}. Likewise, in the second part, there is an anomalous excess between 2003 and 2008, in the declining phase of cycle 23. This seems to be a temporary upward fluctuation coming on top of the slight overall excess marking all measurements after 1983. This feature can also be recognized in the ratios with Wolf numbers from stable individual stations (not illustrated here). 
	
We must stress here that neither the Z\"{u}rich data nor Locarno data played a role in this determination. Therefore, the fact that the transition found in $\rm F_{10.7}$ happens near 1981, the year of the Z\"{u}rich to Locarno transition, is unrelated to this event although the time of this jump roughly coincides. As this 1983-2008 excess in $\rm F_{10.7}$ is found versus multiple independent series, we thus conclude that it is due to a specific feature in the $\rm F_{10.7}$ time series.  This will require deeper investigations to find out if it can be explained by a technical element in the measurements or if it reflects a true but temporary change in solar emission processes.  

\subsection{Inhomogeneities in $\rm R_a $}  \label{SS-Ra}
Soon after the introduction of the American SN $\rm R_a$, large deviations were identified between $\rm R_a$ and the Z\"{u}rich SN. The reasons were only identified in the late 1990s \citep{Schaefer_1997a, Schaefer_1997b, Hossfield_2002}, when methodological flaws were found in the processing of the AAVSO stations. The method was then corrected \citep{Coffey_etal_1999} and the processing was further improved after 2008.

\begin{figure} 
	\centerline{\includegraphics[width=1.0\textwidth,clip=true,trim= 15 0 5 0,clip=true]{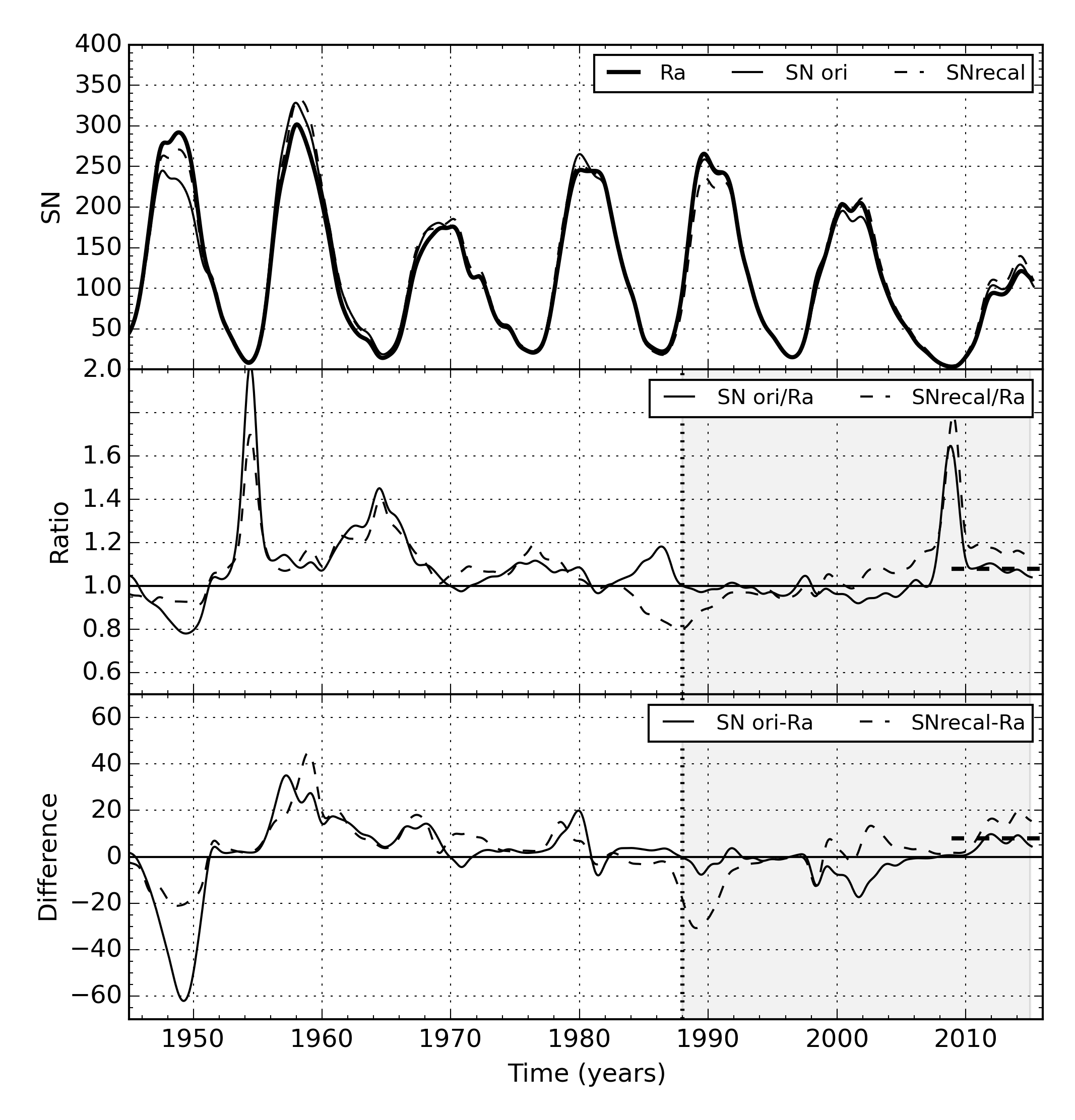}}
	\caption{Comparison of American SN $\rm R_a$ from AAVSO (thick solid line) respectively with the original SN (solid line) and with our re-calibrated SN (dashed line). The middle and lower panels show the ratios and differences respectively. No errors are given here, as the $\rm R_a$ series does not provide error bars. The ratios and differences show large variations until the mid-1980s, which were previously identified and attributed to inhomogeneities in the early $\rm R_a$ series. After 1988 (shaded interval), $\rm R_a$ shows a much closer agreement with the original SN series but surprisingly, differs from our new recalibrated SN by drifts that are very similar to the k correction derived for the 1981-2015 interval (Fig. \ref{Fig-compark}), appearing here in reverse due to the inverse ratio. We also note that both ratios experience a jump to higher values after 2010 (horizontal dashed line), suggesting that recent $\rm R_a$ values are about 10\% below the scale of the preceding time interval.} 
	\label{Fig-compRa}
\end{figure}

Now, comparing our new reconstructed SN to the $\rm R_a$ (Fig. \ref{Fig-compRa}), we recognize the large initial defects until 1970. Thereafter, there is a much better agreement with the Z\"{u}rich SN. Now, what comes as a full surprise is that this good agreement persists after 1981, although we now know that the Brussels-Locarno SN was affected by large variable drifts. Now, when comparing $\rm R_a$ with the new corrected SN series, the ratio shows strong variations that are very similar to the Locarno drift (see figures \ref{Fig-compark} and \ref{Fig-compavgk1945}). 

This strongly suggests that $\rm R_a$ actually used the official international SN as reference and started to track its changing scale. The curves indicate that this cross-dependency started sharply in 1986. Therefore, it seems that a methodological change occurred well before the documented correction of the $\rm R_a$ method in 1999. Unfortunately, the processing for that early epoch is scarcely documented and original raw data from stations were not preserved. Consequently, we must conclude that the existing $\rm R_a$ series is not reliable until the early 2000. Over past years a large effort has been devoted to ensure the stability of the present and future values of $\rm R_a$. However, the comparison in figure \ref{Fig-compRa} also indicates that since the last cycle 23-24 minimum, the values remain systematically too low by a factor $0.901 \pm 0.008$, apparently propagating the negative bias of the second part of cycle 23 up to the present.

\begin{figure} 
	\centerline{\includegraphics[width=0.95\textwidth,clip=true,trim= 15 0 5 0,clip=true]{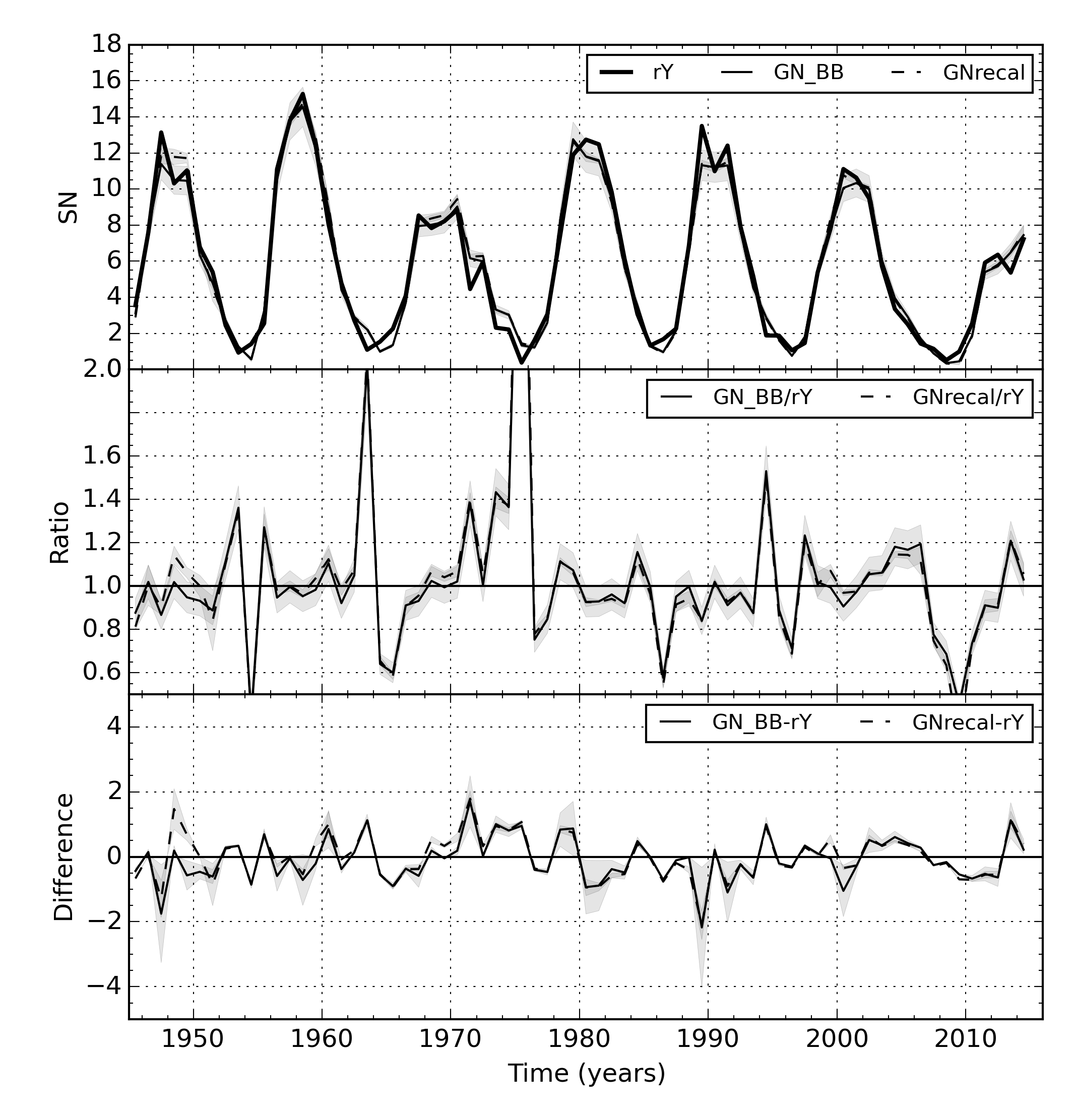}}
	\caption{Comparison of the rY diurnal range of the geomagnetic East component (thick solid line; proxy based on formula \ref{EQ-proxy-rY}) with our reconstructed GN (dashed line) and the new ``backbone'' GN series (solid line), with the corresponding ratios and differences in the middle and lower panels. All three series agree rather well, although the larger uncertainties in the rY series lead to large random fluctuations of the ratio and differences. Still, through an independent reference, this comparison indicates that the scale of our reconstructed GN series remained constant over the entire interval, thus excluding any artificial slow trend in the new GN series.} 
	\label{Fig-compsngn-rY}
\end{figure}

\subsection{Comparisons with the rY diurnal index}  \label{SS-comprY}
\citet{Svalgaard_2010, Svalgaard_2015b} showed a close long-term correlation between the $\rm F_{10.7}$ radio flux and yearly mean diurnal range of the geomagnetic East component rY (a tracer of the ionospheric response to solar UV irradiance) and a similar agreement with the Group Number. As we now have two new reliable reconstructions both of the SN and of the GN, we also checked the comparison with rY in figure \ref{Fig-compsngn-rY}, using as rY-based proxy, the formula:
\begin{equation}
  G_N(rY)= 0.467 \times (rY - 32.67)   \label{EQ-proxy-rY}
\end{equation}
 
The comparison confirms the absence of overall trend in the ratio but due to the higher noise level in the rY series (standard deviation in GN of 0.60, independent of the level of activity) and the coarse yearly sampling, no other more detailed conclusion can be drawn from this comparison.

\section{Conclusions} \label{S-Concl}
The new approach adopted in the present analysis leading to a multi-station average SN series, confirmed both the amplitude and the detailed shape of the time-variable drift of the Locarno pilot station. It also led to new validations and revealed new properties of the SN and GN series. We now have obtained a stable reconstruction of the SN over the period 1981-2015 and thanks to the longer 1945-2015 backbone, we now have a reference allowing to tie this last part of the SN series to the entire Z\"{u}rich series preceding the 1981 transition. In addition, the recalculation of the new SN values allowed us to add standard errors on the resulting average values, based on the actual statistical dispersion of simultaneous counts from multiple stations, an important information that was never included in the original version of the SN series. 

Moreover, regarding the absolute scaling, thanks to the uniformity of the corrected series, the ratios between series over long overlapping periods of up to 30 years have become much more constant, allowing to derive average scaling ratios with accuracy now close to 1 \%. This will thus lead to an improved long-term stability when assembling all parts of the new SN series into the final end-to-end reconstruction over 1749--2015 (Paper II). 

\subsection{New diagnostics}
By exploiting the elimination of large past inhomogeneities, we could establish that the mean relation between the SN and GN is non-linear and can be fitted by a quadratic law that is stable and remains valid over the entire 65 years of the study. We thus conclude that almost all variations in the SN/GN ratio can be explained by the same model and thus that the variations of the past 65 years do not require any change of regime in the solar activity, and in its source, the solar dynamo. The only notable exception is cycle 24, which seems to deviate from earlier cycles and is marked by a definite deficit in the average number of spots per groups. However, our study indicates that if this marks a transition, it occurred only recently, during the cycle 23--24 minimum, and not as part of a progressive decline starting in cycle 23 or earlier. Finding this seemingly universal relation between the GN and SN now raises new questions. Is this relation also valid for the entire early part of the SN and GN series, before 1945? Is the change in the average number of spots per group due to a true excess of spots for high activity levels? Or can it be caused by systematic biases in the visual splitting of groups by observers as suggested by \citet{Svalgaard_2015b}, or by a combination of both effects? Those questions will be the topics of future research and publications.

When examining more closely the differences between the new and the original SN series, we can also observe that the original SN systematically underestimated the second peak for solar cycles 21, 22 and 23. For cycles 22 and 23, it leads to more equal peaks, without changing the absolute maximum of the cycle, while for cycle 23, the maximum moves to the second peak in November 2001 instead of July 2000 originally. It thus now agrees with the shape of maxima found in most other fluxes and indices of solar activity: total solar irradiance, MgII, Ly$\alpha$, total sunspot magnetic flux \citep[see][]{Thuillier_etal_2012, Frohlich_2013, Pevtsov_etal_2014, Snow_etal_2014,  Yeo_etal_2014}. The correction thus also solves more detailed and recent discrepancies between the SN and modern indicators available only over the last 3 or 4 solar cycles. 

This new SN/GN relation illustrates just one of many new studies that can be undertaken on the base of strongly improved SN and GN series. Given their stability and mutual consistency,  those new reference series can also help revisiting the stability of other solar indices. Here, we just considered three immediate equivalents of the SN and we found that the $\rm F_{10.7}$ and $\rm R_a$ series contain scale changes, of which some were suspected but are now clearly revealed when referred to the newly reconstructed SN and GN series.

\subsection{The future SN series}
Many of the methodological improvements developed for this analysis can be exploited to implement a new improved processing of the current and future observations from the worldwide SILSO network. Our results also indicate that the operational processing of new data must be improved in order to avoid or at least mitigate past inaccuracies found in the original SN series. However, for different reasons, the current analysis cannot be directly transposed for this operational purpose. 

Indeed, in the retrospective analysis, we could use the whole history of each station when selecting or eliminating the stations. At a collective level, we could also verify that the network-wide average was not drifting due to some unsuspected global influence. Such a verification is essential when not using a single pilot observatory carefully following pre-defined procedures and rules like the Specola station in Locarno. In this case, we must indeed envisage the possibility of a collective evolution of observing practices caused by external global factors like the technological evolution of commercial astronomical equipment, the emergence of fast Internet communication or the availability of real-time solar images from many sources. When working on archived series, such a verification can be done thanks to multiple comparisons with individual stations and with several alternate solar indices over wide time intervals. Finally, over most of the series, the time averages or running means were derived over a symmetrical window around the time of the reference data point, without edge effects. By contrast, when processing new data to extend the time series, only past information is available and consequently, time averages are truncated for the most recent dates. Moreover, cross-validation with other indices is often only possible after some delay, when enough data have accumulated to derive valid statistics. Therefore, a different processing strategy must be adopted, though still drawing on the methods developed here.

\begin{figure} 
	\centerline{\includegraphics[width=1.0\textwidth,clip=true,trim= 15 0 5 0,clip=true]{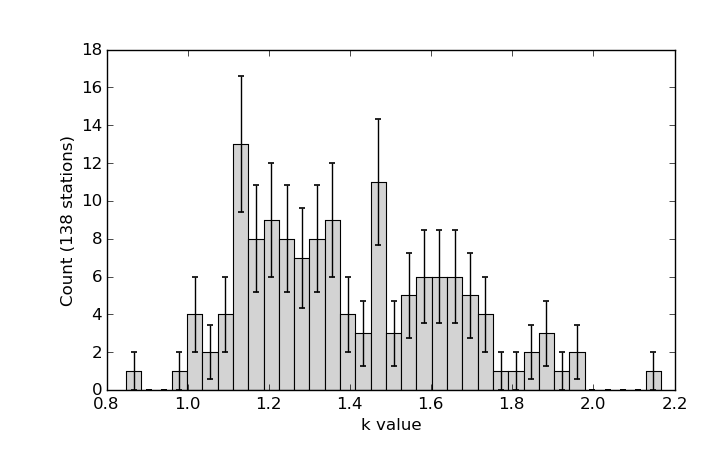}}
	\caption{Histogram of k coefficients for 138 valid stations in the SILSO network, with an activity period longer than 11 years. The mean k ratios were computed over the entire 1981-2015 interval relative to the new recalibrated SN number. Thanks to the homogeneity of the new SN, the individual k ratios are now mostly constant over the entire time interval. Error bars give the uncertainty on the counts in each bin (Poissonian statistics).  The distribution is rather broad (factor 2) and features several narrow and broad peaks, marking subpopulations among the stations. Except for a few cases, most stations have a k larger than one, and are thus counting less spots than the Locarno pilot station (k=1), in part because SILSO stations do not weight their counts according to spot size, by contrast with the Locarno station. } 
	\label{Fig-histok-sn}
\end{figure}

One aspect of the operational SN production that will directly benefit from our analysis is the selection or rejection of valid stations and the determination of their global statistical properties. Now that we can refer all stations to a common stable reference, their corresponding k ratio is also stable over long durations, often the full length of the observations. Therefore, we can now compute a single average k value for each station and produce an histogram for all stations (Fig. \ref{Fig-histok-sn}). We can now see that the distribution of k values has a multi-peaked distribution. The same main peaks are present in the histogram resulting from the monthly and yearly means, and are thus clearly not due to random statistical fluctuations (Poissonian statistics: standard deviations given by the error bars). The SILSO network thus contains several sub-populations of stations giving similar sunspot counts. 

\begin{figure} 
	\centerline{\includegraphics[width=1.0\textwidth,clip=true,trim= 15 0 5 0,clip=true]{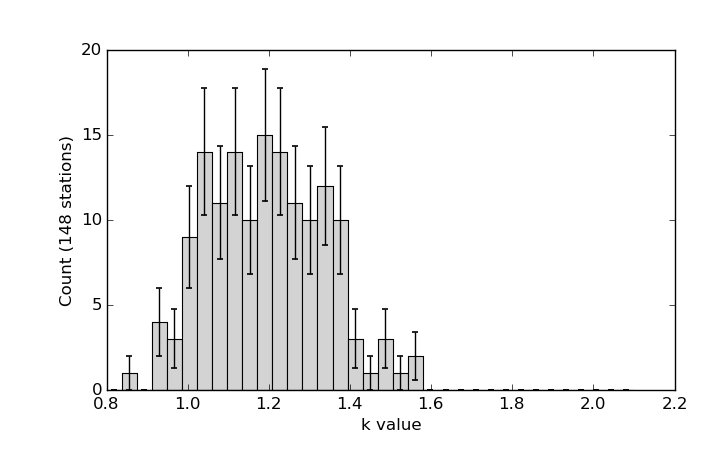}}
		\caption{Histogram of k coefficients for 148 valid stations in the SILSO network, with an activity period longer than 11 years. This histogram is equivalent to the plot of figure \ref{Fig-histok-sn} but for the Group Numbers. Here, the distribution is much more compact  (factor 1.5) and closer to unity. It is also smoother without distinct peaks for specific k values. However, here again, most stations count less groups than the Specola--Locarno reference station (k=1). } 
		\label{Fig-histok-gn}
\end{figure}

The main peak at 1.15 below the Locarno scale probably includes stations counting as many spots as Locarno but without using a weighting according to size, which causes an excess of the order of 18 \% \citep{Clette_etal_2014}. The properties of stations at k=1.30, 1.45 and 1.65 are less clear and deserve further study. The last stations at 1.19 give Wolf numbers that are 0.6 times lower than the numbers from stations in the main 1.15 peak, i.e. matching the original numbers by Wolf himself using his small portable telescopes (4\,cm apertures equivalent to typical modern binoculars). The discovery of those clear sub-populations must definitely be taken into account for implementing the statistical treatment leading to the sunspot number. By contrast, the histogram of k coefficients for the Group Numbers (Fig. \ref{Fig-histok-gn}) gives a fairly smooth distribution, without significant concentration around specific values, thus indicating that the above sub-groups of stations differ mainly by the spot-count term in their Wolf numbers (Equ. \ref{EQ_wolfnum}). 

Finally, in the course of our analyses, we also learned that although many stations proved to be stable references in general, all those stations experience occasional changes in their personal k coefficient, producing dips or peaks often for limited durations. These can be typically caused by a change of observer in professional observatories or by the aging of individual observers at the end of their observing career (degrading eyesight). Therefore, we conclude that it is impossible to build a fully stable reference by using a single pilot station, as was done until now. Moreover, relying on a single station always induces a risk, as any interruption or termination of the observations implies the loss of the long-term scale. 

Now that the past series has been re-calibrated, our next goal will thus be the construction of a multi-station reference, by selecting a core group of stable stations from the SILSO network. With multiple stations, continuous cross-comparisons become possible and will allow detecting any abnormal deviation occurring only for a single member of the core group. The scale of the resulting sunspot numbers will then be much less sensitive to accidental anomalies at a single station. For the selection of the candidate stations, we can now directly use the ranking according to linear correlation and residual dispersions resulting from the recalibration analysis presented here. 

Overall, the present re-calibration was essential for consolidating the final link between the present and the entire sunspot record of past centuries and simultaneously, it opens the way to a new improved continuation of the re-calibrated series for years to come.


\begin{acks}
F.Clette and L. Lef\`{e}vre would like to acknowledge financial support from the Belgian Solar-Terrestrial Center of Excellence (STCE; \url{ http://www.stce.be}). Part of this work was developed in the framework of the SOLID project (EU $\rm 7^{th}$ Framework Program, SPACE collaborative projects, \url{http://projects.pmodwrc.ch/solid/}) and of the TOSCA project (ESSEM COST action ES1005 of the European Union; \url{http://lpcs2e.cnrs-orleans.fr/~ddwit/TOSCA/Home.html}).
\end{acks}

%
%
\bibliographystyle{spr-mp-sola}
\bibliography{BibClette}  
%
%
%
%

\end{article} 
\end{document}